\documentclass[preprint,sort&compress]{elsarticle}

\usepackage{lineno}

\usepackage{hyperref}
\usepackage{graphicx}
\usepackage{siunitx}
\usepackage{multirow}

\usepackage[left=3.5cm,right=3.5cm,top=1.6cm,bottom=1.6cm,includeheadfoot]{geometry}

\journal{NIM A}

\newcommand*{\hebar}{\overline{\mathrm{He}}}
\newcommand*{\hebart}{^3\overline{\mathrm{He}}}
\newcommand*{\hebarf}{^4\overline{\mathrm{He}}}

\mathchardef\mhyphen="2D

\begin{document}

\begin{frontmatter}

\title{AMS-100:
The Next Generation Magnetic Spectrometer in Space -- 
An International Science Platform for Physics and Astrophysics at Lagrange Point~2}

\author[ac1]{S.~Schael\corref{ca}}
\cortext[ca]{Corresponding author}
\ead{schael@physik.rwth-aachen.de}
\author[mmi]{A.~Atanasyan}
\author[ciemat]{J.~Berdugo}
\author[ac3]{T.~Bretz}
\author[fh]{M.~Czupalla}
\author[fh]{B.~Dachwald}
\author[hi]{P.~von~Doetinchem}
\author[perugia]{M.~Duranti}
\author[ac1]{H.~Gast}
\author[ac1]{W.~Karpinski}
\author[ac1]{T.~Kirn}
\author[ac1]{K.~L\"{u}belsmeyer}
\author[ciemat]{C.~Ma\~{n}a}
\author[siena]{P.~S.~Marrocchesi}
\author[ttk]{P.~Mertsch}
\author[wwh]{I.~V.~Moskalenko}
\author[sla]{T.~Schervan}
\author[mmi]{M.~Schluse}
\author[sla]{K.-U.~Schr\"{o}der}
\author[ac1]{A.~Schultz~von~Dratzig}
\author[dqmp]{C.~Senatore}
\author[fh]{L.~Spies}
\author[uc]{S.~P.~Wakely}
\author[ac1]{M.~Wlochal}
\author[psi]{D.~Uglietti}
\author[sla]{J.~Zimmermann}

\address[ac1]{I.~Physikalisches Institut, RWTH Aachen University,
  Sommerfeldstr.~14, 52074 Aachen, Germany}
\address[mmi]{Institut f\"{u}r Mensch-Maschine-Interaktion, RWTH Aachen
  University, Ahornstr.~55, 52074 Aachen, Germany}
\address[ciemat]{Centro de Investigaciones Energ\'{e}ticas, Medioambientales y
  Tecnol\'{o}gicas (CIEMAT), Av.~Complutense 40, 28040 Madrid, Spain}
\address[ac3]{III.~Physikalisches Institut A, RWTH Aachen University,
  Sommerfeldstr.~14, 52074 Aachen, Germany}
\address[fh]{Fachbereich Luft- und Raumfahrttechnik, Fachhochschule
  Aachen, Hohenstaufenallee 6, 52064 Aachen, Germany}
\address[hi]{Physics and Astronomy Department, University of Hawaii, Honolulu, HI, 96822, U.S.A.}
\address[perugia]{INFN Sezione di Perugia, 06100 Perugia, Italy}
\address[siena]{Department of Physical Sciences, Earth and Environment, University of Siena and INFN Sezione di Pisa, 53100 Siena, Italy}
\address[ttk]{Institut f\"{u}r Theoretische Teilchenphysik und Kosmologie, RWTH Aachen University,
  Sommerfeldstr.~14, 52074 Aachen, Germany}
\address[wwh]{W.W.~Hansen Experimental Physics Laboratory, Kavli
  Institute for Particle Astrophysics and Cosmology, Department of
  Physics and SLAC National Accelerator Laboratory, Stanford
  University, Stanford, CA, 94305, U.S.A.}
\address[sla]{Institut f\"{u}r Strukturmechanik und Leichtbau, RWTH Aachen
  University, W\"{u}llnerstr.~7, 52062 Aachen, Germany}
\address[dqmp]{Department of Quantum Matter Physics, Universit\'{e} de
  Gen\`{e}ve, 24 Quai Ernest-Ansermet, 1211 Geneva, Switzerland}
\address[uc]{Enrico Fermi Institute, University of Chicago, Chicago, IL, 60637, U.S.A.}
\address[psi]{Ecole Polytechnique F\'{e}d\'{e}rale de Lausanne (EPFL), Swiss Plasma Center (SPC), 5232 Villigen PSI, Switzerland}

\begin{abstract}
  The next generation magnetic spectrometer in space, AMS-100, is
  designed to have a geometrical acceptance of \SI{100}{m^2.sr} and to
  be operated for at least ten years at the Sun-Earth Lagrange
  Point~2. Compared to existing experiments, it will improve the
  sensitivity for the observation of new phenomena in cosmic rays, and
  in particular in cosmic antimatter, by at least a factor of
  1000. The magnet design is based on high temperature superconductor
  tapes, which allow the construction of a thin solenoid with a
  homogeneous magnetic field of \SI{1}{Tesla} inside. The inner volume
  is instrumented with a silicon tracker reaching a maximum detectable
  rigidity of \SI{100}{TV} and a calorimeter system that is 70
  radiation lengths deep, equivalent to four nuclear interaction
  lengths, which extends the energy reach for cosmic-ray nuclei up to
  the PeV scale, i.e.~beyond the cosmic-ray knee. Covering most of
  the sky continuously, AMS-100 will detect high-energy gamma rays in
  the calorimeter system and by pair conversion in the thin solenoid,
  reconstructed with excellent angular resolution in the silicon
  tracker.
\end{abstract}


\end{frontmatter}



\section{Introduction} 
\begin{figure*}[t]
  \centering
  \includegraphics[width=\textwidth]{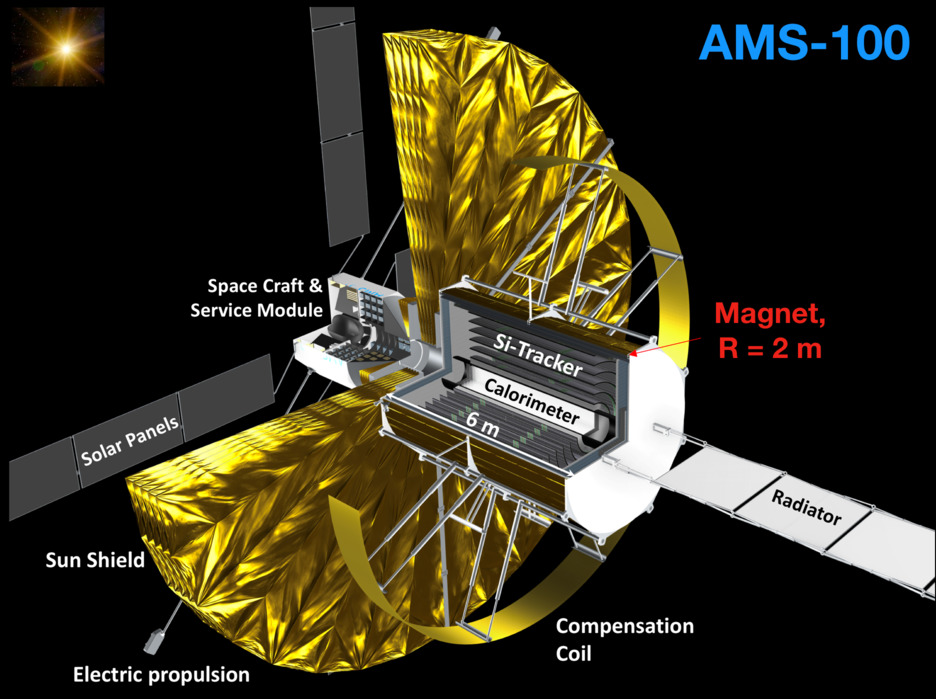}
  \caption{AMS-100 detector concept.
  \label{fig:concept}}
\end{figure*}

{\bf A} {\bf M}agnetic {\bf S}pectrometer with
a geometrical acceptance of {\bf 100} \si{m^2.sr}, {\bf AMS-100}, is a
major new space mission which addresses a number of key science
questions in multi-messenger astrophysics, cosmic-ray physics and
particle physics (Fig.~\ref{fig:concept}). Several of these questions
have emerged in the last decade, as a result of the tremendous success
of recent space missions, such as PAMELA~\cite{Picozza2007},
Fermi-LAT~\cite{Atwood2009}, AMS-02~\cite{Kounine2012},
CALET~\cite{CALET}, and DAMPE~\cite{Gargano_2017}. In particular, the
magnetic spectrometer AMS-02 has revealed several unexpected new
features in the cosmic-ray
matter~\cite{PhysRevLett.121.051103,Aguilar2019a} and antimatter
fluxes~\cite{Aguilar2016,Aguilar2019} that have challenged much of our
traditional understanding of particle astrophysics, across a range of
topics such as the nature of dark matter and the origin and
propagation of cosmic rays. Direct measurements of cosmic rays provide
important constraints to trace the structure of the Galaxy, and to
search for signatures of new
physics~\cite{Klasen:2015uma,J_hannesson_2018,Gabici:2019jvz}. Even
more important could be the observation of $\hebar$ candidate events
in cosmic rays~\cite{hebar}, which could have profound implications
for understanding the origin of the matter-antimatter asymmetry of the
universe.

These questions cannot be addressed by calorimeter-based instruments
in space, which, in the absence of magnetic deflection, can measure
neither the charge sign nor the mass of the incoming particles.
Therefore we believe that ground-breaking progress for fundamental
physics requires a next generation magnetic spectrometer in space. Due
to the strong dependence of the cosmic-ray flux $\Phi$ on energy $E$,
approximated by $\Phi\propto{}E^{-3}$, every increase in energy reach
by a factor of 10 requires an increase in geometrical acceptance by a
factor of 1000.

Simply scaling the dimensions for the telescope-like geometries of
PAMELA or AMS-02 would not allow significantly enhancing the
geometrical acceptance and the energy reach at the same time.
Increasing the height would enhance the energy reach but reduce the
acceptance. Increasing the diameter would enhance the acceptance but
reduce the magnetic field for a fixed magnet wall thickness and hence
the energy reach. This dilemma can only be overcome by moving to a
different detector geometry. A possible solution has been pioneered
successfully by the BESS experiment~\cite{YAMAMOTO199475} with a thin
solenoid. The key here is the fact that the central magnetic field for
a long solenoid only depends on the number of turns, the current and
the length, but not on the radius. Therefore, for a solenoid of given
wall thickness and instrumented with a tracking detector on the inside
like a classical collider experiment, both the geometrical acceptance
and the maximum detectable rigidity (MDR, defined by $\Delta_R/R=1$,
where $\Delta_R$ is the uncertainty of the rigidity measurement)
increase quadratically with the radius if the diameter-to-length ratio
stays constant. If placed far away from Earth, such an instrument has
an angular acceptance of up to $4\pi$~steradian due to its rotational
symmetry, superior to any telescope-like geometry.

The instrument described in this article will explore uncharted
territory in precision cosmic-ray physics by employing a suite of
sophisticated detector systems designed to improve on existing
instrumentation in both precision and in energy reach.  The key
element of the instrument is a thin, large-volume high temperature
superconducting (HTS) solenoid magnet which creates a homogeneous
magnetic field of \SI{1}{Tesla} in the tracking volume. It is cooled
passively to \SIrange{50}{60}{K}. An expandable compensation coil
balances the magnetic moment of the solenoid and allows the attitude
control of the instrument within the heliospheric magnetic field.
Combining this powerful solenoid with proven tracking technologies and
innovative ``cubic'' calorimetry designs, the spectrometer will
achieve an \hypertarget{q:mdr}{MDR} of \SI{100}{TV}, with an effective
acceptance of \SI{100}{m^2.sr}. The central calorimeter has a depth of
70 radiation lengths ($X_0$), or 4 nuclear interaction lengths
($\lambda_I$). This instrumentation will allow probing, with high
statistical power and high precision, the positron and electron
spectra to \SI{10}{TeV}, the antiproton spectrum to \SI{10}{TV}, and
the nuclear cosmic-ray component to
\SI[retain-unity-mantissa=false]{1e16}{eV}, past the cosmic-ray knee.

For the first time, this instrument will have the acceptance and
resolution to probe the cosmic-ray antideuteron spectrum with high
precision. AMS-100 will vastly expand our sensitivity to heavy cosmic
antimatter ($Z\leq{}-2$).  Covering most of the sky continuously,
AMS-100 will provide high-resolution survey measurements of $\gamma$
rays to energies beyond the TeV scale, with an angular resolution of
\ang{;;4} at \SI{1}{TeV} and \ang[angle-symbol-over-decimal]{;;0.4} at
\SI{10}{TeV}, comparable to X-ray telescopes~\cite{xmm}.

The instrument will be installed on a spacecraft and operated
for at least \hypertarget{q:duration}{ten years} at the Sun-Earth
Lagrange Point~2 (L2). This positioning is necessary to
create a stable cold environment for the operation of the HTS
magnet. In a low-Earth orbit, the interaction of the residual
magnetic moment with the geomagnetic field would result in a complete
loss of attitude control. In addition, the shadow of the Earth would
reduce the field of view and the geomagnetic cutoff would limit the
sensitivity to low-energy cosmic antimatter, in particular to
antideuterons from dark matter annihilations.

\begin{figure}
  \centering
  \includegraphics[width=\columnwidth]{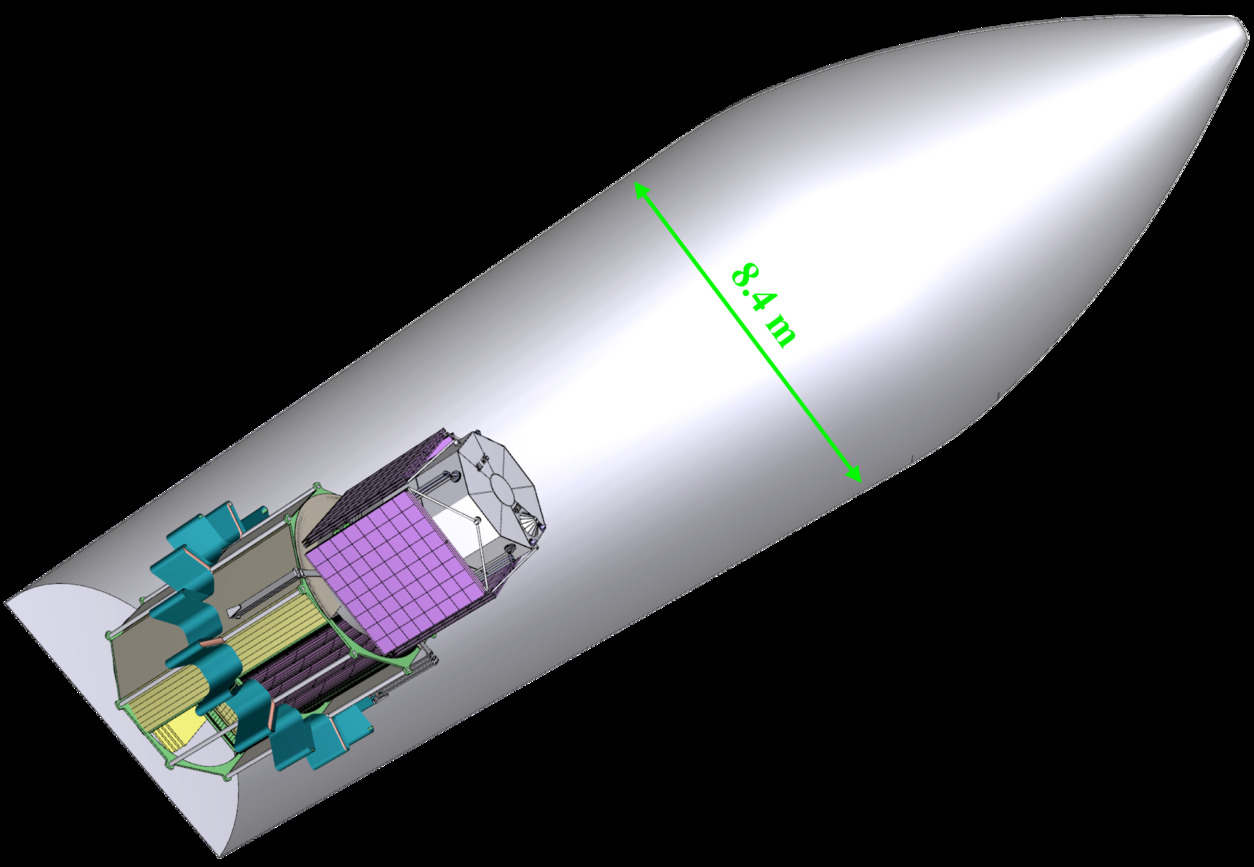}
  \caption[AMS-100 launch configuration.]{AMS-100 launch configuration
    in an SLS-Block 2 fairing. The compensation coil, the sunshield,
    the solar cells, and the electric propulsion system are folded
    up. The service module is located at the top for structural
    reasons.
  \label{fig:launch}}
\end{figure}

To fulfill the science requirements, the full payload has a mass of
\SI{40}{tons} and hence requires new heavy-lift launch
capabilities such as NASA's Space Launch System (SLS) or China's Long
March 9 rocket, which are under development for human missions to
Mars. Figure~\ref{fig:launch} illustrates the launch configuration in
an SLS fairing.

A plausible timeline for instrument definition, design, development,
and testing would target a launch date in 2039, though this requires
an early commitment from the agencies and the community to perform the
necessary R\&D tasks. This will include some level of underlying
technology development, as well as a pathfinder mission to test the
high temperature superconducting solenoid magnet system at L2.

The purpose of this article is a description of the general detector
concept. Several publications will follow describing the magnet
system, the event trigger and DAQ system, the structural and thermal
concept, the service module, the individual sub-detector systems, the
pathfinder mission, and the physics program in detail.

\section{AMS-100 Magnet System}
The geometrical acceptance of \SI{100}{m^2.sr} defines the dimensions
of the \SI{3}{mm} thin main solenoid. It has a length of \SI{6}{m} and
a diameter of \SI{4}{m} (Fig.~\ref{fig:concept}) and creates a central
magnetic field of \SI{1}{Tesla} along the $z$-axis. As the magnet will
be operated at \SIrange{50}{60}{K}, the only option is to construct it
from second-generation rare-earth barium copper oxide (REBCO) high
temperature superconducting
tapes~\cite{Selvamanickam2009,Senatore_2014}. These HTS tapes have a
typical thickness of $\sim$\SI{0.1}{mm} and can carry high current
densities even at field strengths of \SI{30}{T} \cite{refId0} and
tolerate severe mechanical stresses \cite{Barth_2015,Ilin_2015} thanks
to a \SIrange{30}{100}{\um} thick Hastelloy substrate. Today, REBCO
tapes are available in high lengths~\cite{Fujikura2019}, and are
studied in several research projects. A typically \SI{20}{\um}
thick copper stabilizer completes the HTS tapes (for more details, see
for example Refs.~\cite{SuperPower,Fujikura,Zhao_2019}) which can be
easily soldered for joints. It has been shown in Ref.~\cite{6084824}
that increasing the stabilizer thickness aids in reducing the magnet
temperature at a quench. For the AMS-100 magnets, we assume that the
copper stabilizer will be replaced by an equivalent aluminum
stabilizer to minimize the material budget~\cite{6983566}.

Quench protection and understanding the dynamics of the quench process
in HTS tapes~\cite{doi:10.1063/1.4954165} are the key for the long
term stable operation of such a magnet in space. As one possible
option, HTS coils can be protected from an irreversible quench by
winding them from tapes without additional
insulation~\cite{5675711,YANAGISAWA201440,Suetomi_2019}, thus allowing
the current to flow in the radial direction in case of a thermal
runaway.

Generally REBCO tapes are available in piece lengths of
\SIrange{300}{500}{m} with joint resistances of less than
\SI{20}{n\ohm}~\cite{LALITHA20177}.  For a \SI{450}{m} long REBCO tape
at $T=\SI{50}{\kelvin}$ and a magnetic field of \SI{1}{T}, a critical
current of $I_c=$\SI{1000}{A/cm \mhyphen wide}, equivalent to
$I_c=$\SI{1200}{A} for \SI{1.2}{cm} wide tape, has been reported in
2019~\cite{Fujikura2019}.

\begin{figure}
  \centering
  \includegraphics[width=\columnwidth]{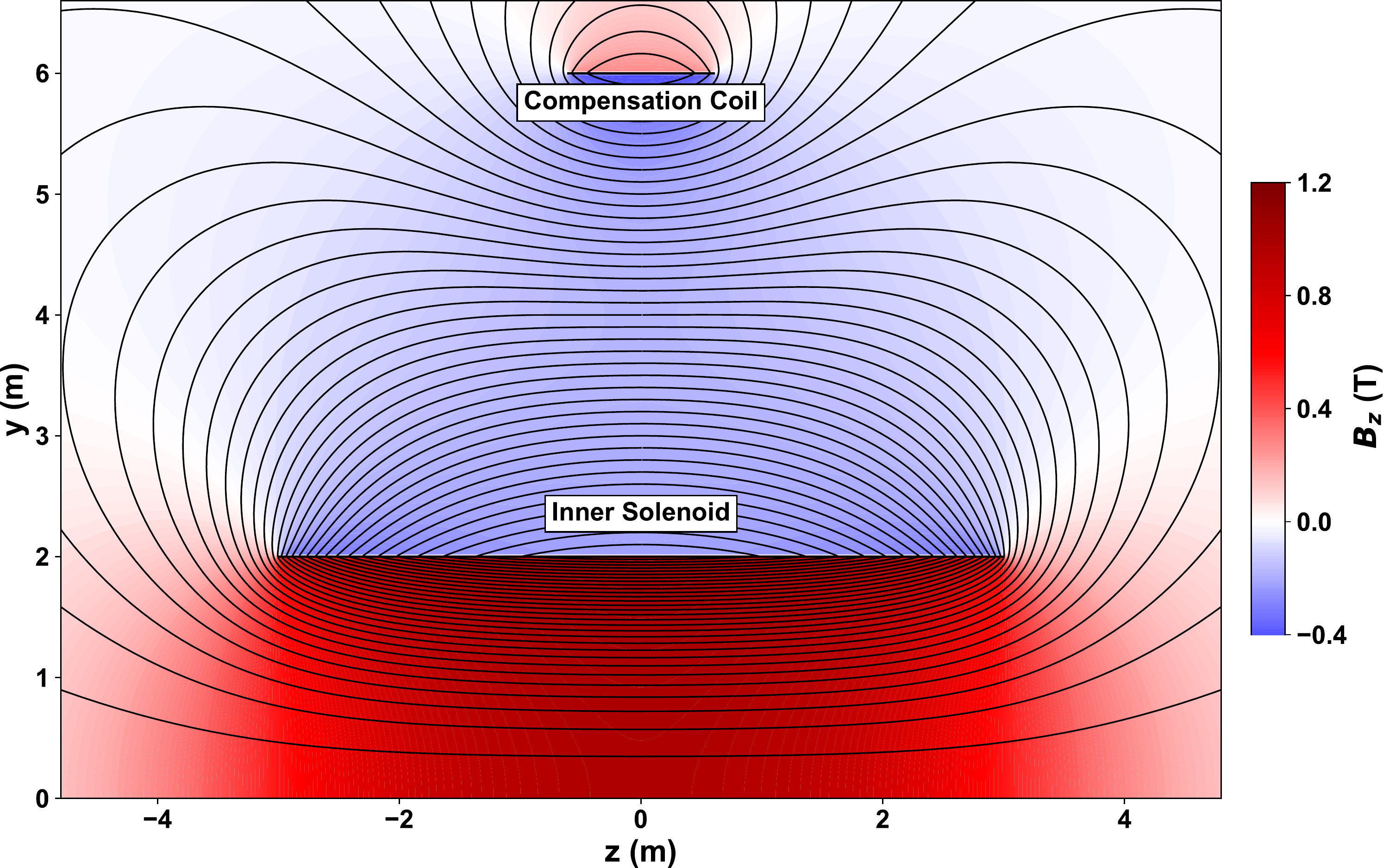}
  \caption[Magnet System.]{Magnetic field lines in the AMS-100 magnet
    system (black) and amplitude of the $z$-component of the magnetic
    field (color map). The compensation coil cancels the magnetic moment of the
    main solenoid, without substantially affecting the magnetic field
    inside the main solenoid.
    \label{fig:magnet}}
\end{figure}

The key parameters of the magnet system for AMS-100 are given in
Table~\ref{tab:magnet}. Progress on the critical current $I_c$ for
REBCO tapes, as expected in the coming years, will proportionally reduce
the number of layers required to obtain a central magnetic field of
\SI{1}{T} and will hence allow reducing the weight and the material
budget of the coils further. The magnetic field is visualized in
Fig.~\ref{fig:magnet}.

\begin{figure*}
  \centering
  \includegraphics[width=\textwidth]{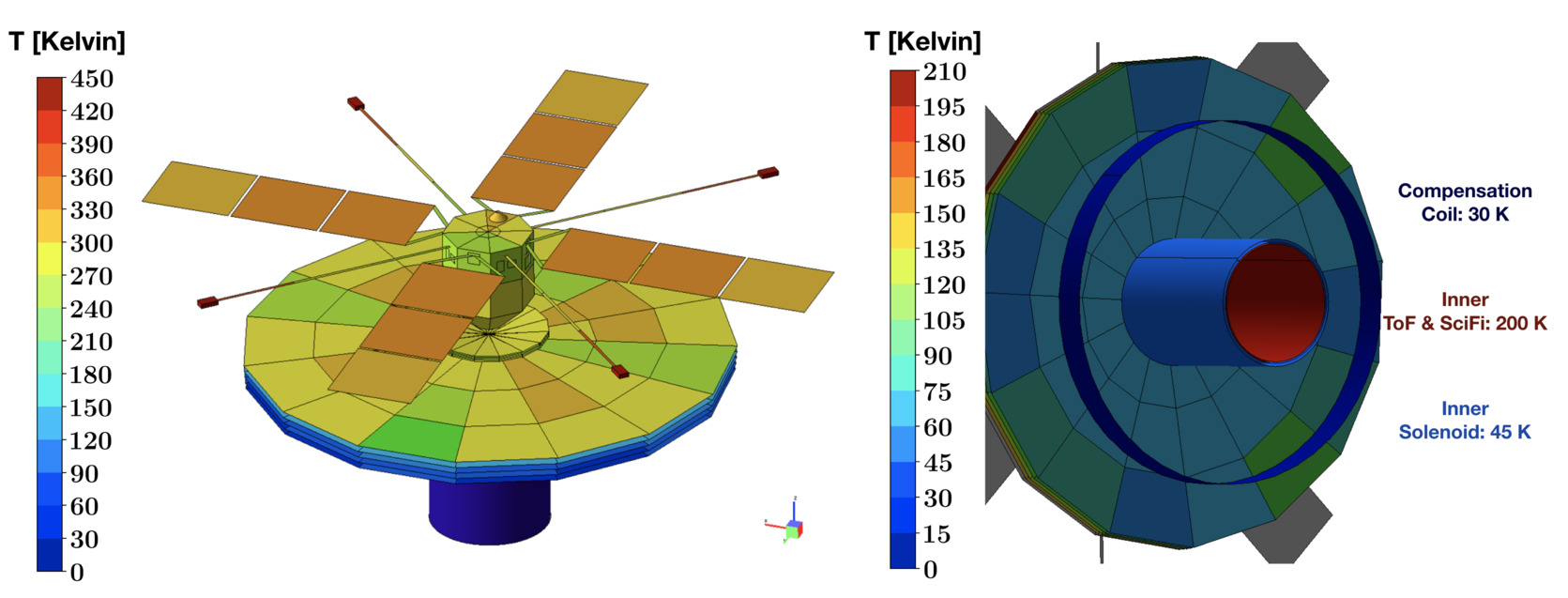}
  \caption[Thermal model.]{Simplified thermal model for AMS-100 taking
    only the radiation between the surfaces, the Sun and deep space
    into account. The color scale indicates temperature in Kelvin. 
    Left: warm side facing the Sun, right: cold side facing deep
    space.
    \label{fig:thermal}}
\end{figure*}

The thin solenoid is cooled by radiation to deep space and operated in
thermal equilibrium at a temperature of \SIrange{50}{60}{\kelvin}
behind a sunshield. A simplified thermal model taking only radiation
into account is shown in Fig.~\ref{fig:thermal}. The main solenoid is
insulated thermally from the other detector components by multi-layer
insulation. The obtained magnet temperatures leave some margin for
conductive thermal loads which have to be taken into account in the
final thermo-mechanical design. Similar to all other detectors inside
the main solenoid, the silicon tracker temperature will be kept
constant at \SI{200}{\kelvin} using a two-phase cooling system or heat
pipes connected to the radiator opposite the sunshield
(Fig.~\ref{fig:concept}). This temperature of \SI{200}{\kelvin} might
have to be adjusted within the overall thermo-mechanical model to
ensure a stable operating temperature for the main solenoid of
\SIrange{50}{60}{\kelvin}. All sub-detector systems are designed to
have a better signal-to-noise ratio at such low temperatures than at
room temperature and first laboratory tests of various detector
components down to liquid nitrogen temperatures have already been
performed successfully at RWTH Aachen~\cite{Erpenbeck}.

\begin{table}
  \centering
  \begin{tabular}{lrr}
                                  & Main                & Compensation \\
                                  & solenoid            & coil \\ \hline
                                  
 Inner radius                    & \SI{   2.0}{m}       & \SI{   6.0}{m}        \\
 Length                          & \SI{   6.0}{m}       & \SI{   1.2}{m}        \\
 Current                         & \SI{   500}{A}       & \SI{  1500}{A}        \\
 Temperature                     & $50\,\mhyphen\,60$~K & $30\,\mhyphen\,40$~K  \\
 HTS tape width                  & \SI{   12}{mm}       & \SI{  12}{mm}         \\
 HTS tape layers                 &    22                &   4                   \\
 $B_z$ at center                 & \SI{   1.0}{T}       & \SI{ -0.06}{T}        \\
 Stored energy                   & \SI{    37}{MJ}      & \SI{   4.5}{MJ}       \\
 Magnetic moment                 & \SI{    70}{MA.m^2}  & \SI{   -70}{MA.m^2}   \\
 Coil thickness                  & \SI{   3.0}{mm}      & \SI{   0.5}{mm}       \\
 Mass                            & \SI{   1.2}{t}       & \SI{  0.13}{t}        \\
 Volume                          & \SI{    75}{m^3}     & \SI{   136}{m^3}      \\
\multirow{2}{*}{Material budget} & $  0.12\,X_0$        & $  0.02\,X_0$         \\
                                 & $ 0.012\,\lambda_I$  & $ 0.002\,\lambda_I$   \\
 Wire length                     & \SI{   150}{km}      & \SI{    15}{km}       \\
 Hoop stress $\sigma_{\theta}$   & \SI{   270}{MPa}     & \SI{   250}{kPa} \\
 $\sigma_{R}$                    & \SI{  -130}{kPa}     & \SI{   -40}{kPa} \\
 $\sigma_{Z}$                    & \SI{  -140}{MPa}     & \SI{   -79}{kPa} \\
 
  \end{tabular}
  \caption{Main parameters of the AMS-100 magnet system. The
    mechanical stresses are denoted by $\sigma$ and calculated
    according to the formulae given by Iwasa~\cite{Iwasa}.
    \label{tab:magnet}}
\end{table}

\begin{figure}
  \centering
  \includegraphics[width=\columnwidth]{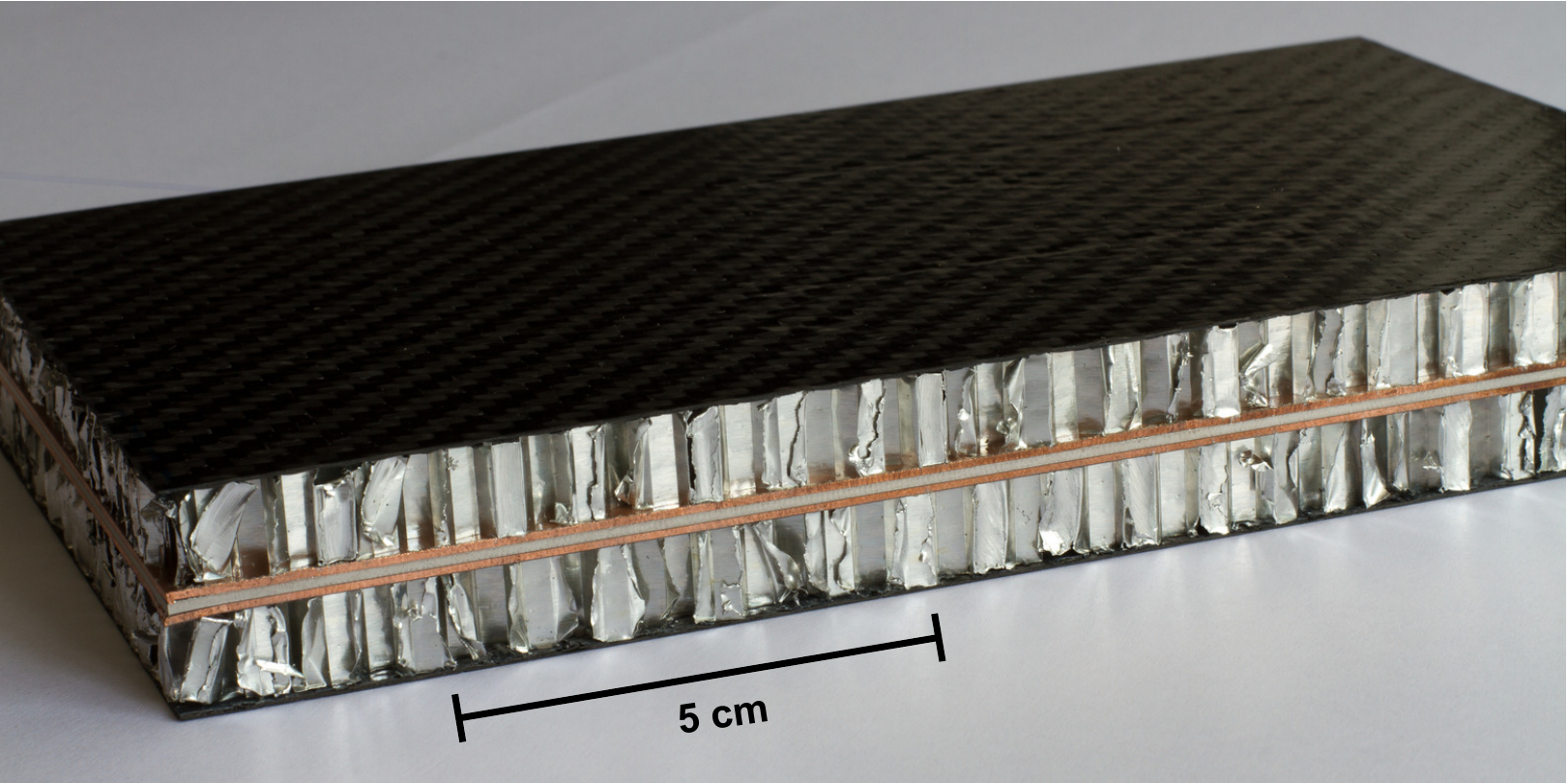}
  \caption{Photograph of a structural test article 
  for the lightweight support of the AMS-100 main solenoid. The
  central layer is mechanically equivalent to the expected magnet.
  \label{fig:mag_support}}
\end{figure}

Particularly for the sensitivity to antimatter in cosmic rays, the
small wall thickness of the main solenoid and its support structure
are of key importance. One option for this that we have
studied in more detail consists of two lightweight aluminum honeycomb
structures with a height of \SI{10}{mm} each that sandwich the magnet
and have carbon fiber face sheets on the outside
(Fig.~\ref{fig:mag_support}). The coil would be assembled on a
temporary support and afterwards the outer honeycomb and carbon fiber
face sheets would be laminated directly onto the outer side of the
magnet. In the next step, the temporary inner support would be removed
and the inner honeycomb and carbon fiber face sheets would be
laminated.  The total material budget of this structure would be
equivalent to a solid aluminum cylinder of \SI{3}{mm} thickness
($0.04\,X_0$). The further optimisation of this lightweight magnet
support structure will have to take all components of the instrument
and the constraints from the thermal model into account.

\begin{figure}
  \centering
  \includegraphics[width=\columnwidth]{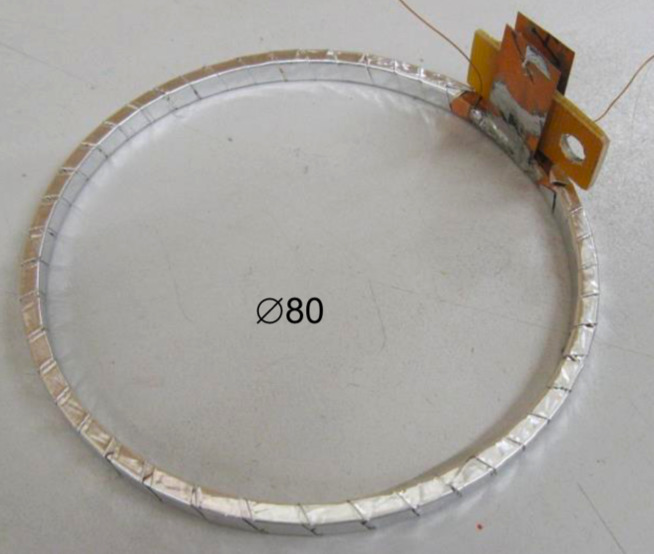}
  \caption{Photograph of a 20 layer HTS test pancake with a diameter
    of \SI{80}{mm}.
  \label{fig:mag_prototype}}
\end{figure}
It has never been demonstrated that a HTS magnet with a lightweight
support structure can be operated in space. In particular, the
vibrations during the rocket launch are a concern. We have therefore
started to produce first prototypes (Fig.~\ref{fig:mag_prototype}) of
thin HTS pancakes to performe space qualification tests including
vibration and thermo-vacuum tests.

\begin{figure*}
  \centering
  \includegraphics[width=\textwidth]{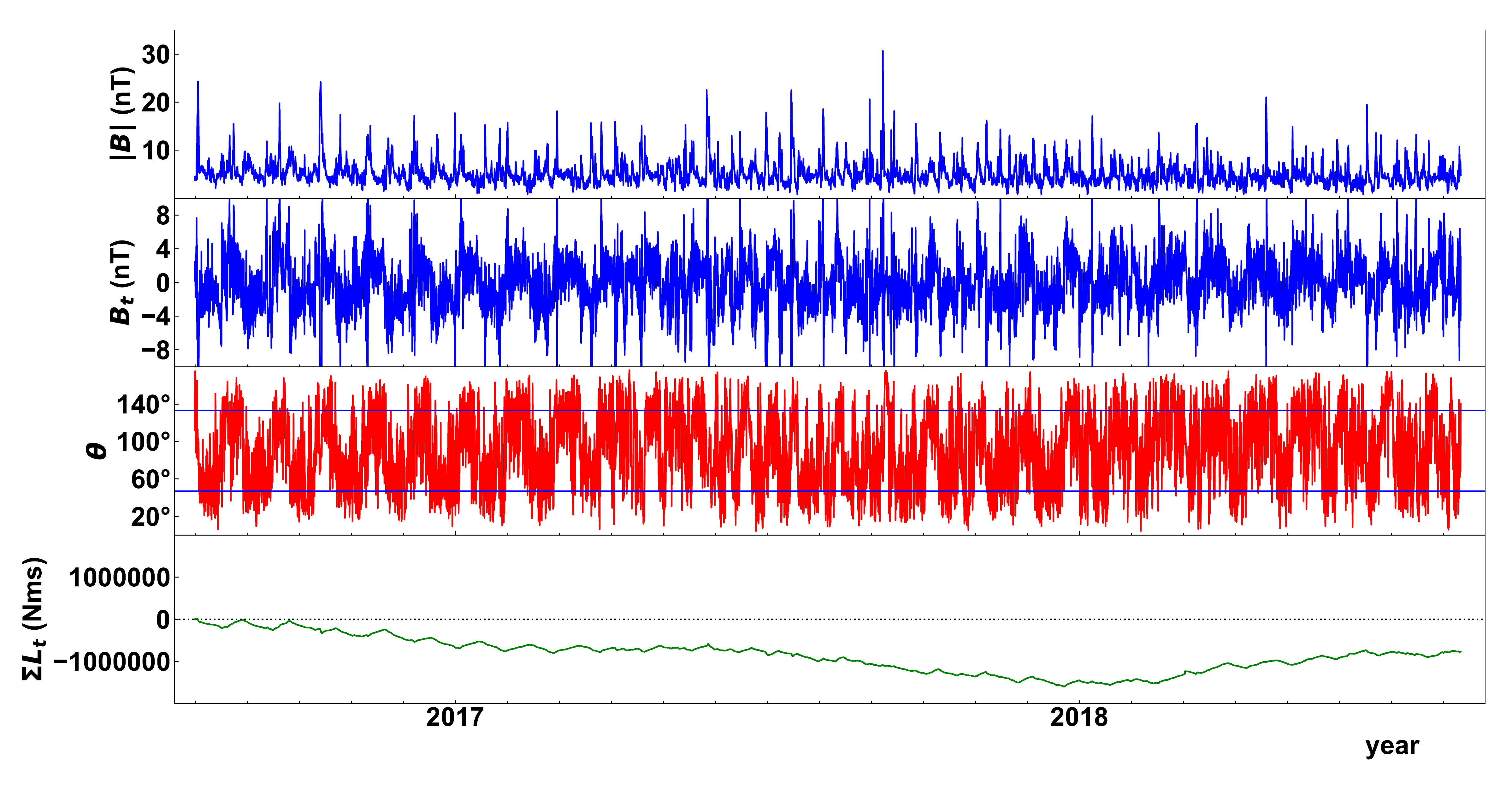}
  \caption[Solar magnetic field at L1.]{Properties of the solar
    magnetic field based on recent measurements by the ACE/MAG
    instrument~\cite{acemag} at Lagrange Point 1 (L1)
    and resulting angular momentum accumulated by the
    main solenoid of AMS-100 without a compensation coil.
    From top to bottom:
    $|\mathbf{B}|$, norm of the interplanetary magnetic field;
    $B_t$, its transverse component relative to the line between the
    Sun and L1;
    $\theta$, the angle between the magnetic field vector and
    this line. The horizontal blue lines mark the values of
    $\theta$ calculated for a simple Parker spiral field
    geometry for the two heliospheric polarities;
    $\sum{}L_t$, angular momentum accumulated by the main solenoid in
    transverse direction.
    \label{fig:solarfield}}
\end{figure*}

For the operation of a large solenoid in deep space, the interaction
with the interplanetary magnetic field (IMF) is a major concern.  The
IMF has a complicated time-dependent structure. Due to the rotation of
the Sun (period of \SI{25.4}{days}), its magnetic field winds up into
a large rotating spiral. The heliospheric magnetic field changes
polarity every $\simeq$11 years~\cite{Parker2001}. It is distorted at the
orbit of AMS-100 around L2 by the geomagnetic field and by solar
flares. Due to the solar wind, the magnetic field at L2 still has an
average strength of \SI{6}{nT}, varying between 0 and \SI{35}{nT}. In
combination with the large magnetic moment of the AMS-100 main
solenoid, this causes an average torque of \SI{0.4}{N.m}. Based on
measurements of the heliospheric magnetic field at Lagrange Point 1,
which is very close to L2 on heliospheric scales, we can derive the
expected angular momentum as a function of time
(Fig.~\ref{fig:solarfield}). Even though the magnetic field reverses
polarity periodically, the accumulated angular momentum reaches a
value on the order of $10^6\,\si{N.m.s}$ over the course of one
year. Such a large angular momentum cannot be balanced by reaction
wheels or control moment gyroscopes. Instead, a compensation coil is
needed with opposite field direction to balance the magnetic dipole
moment of the main solenoid (Fig.~\ref{fig:magnet}).

With a diameter of \SI{12}{m}, the compensation coil has to be an
expandable coil, as has been studied for radiation shielding in space
in~\cite{NASA-HTS}. It will consist of \SI{0.5}{\mm} of HTS tape embedded
and reinforced by \SI{1}{mm} thick kevlar or zylon layers. The support
structure of this coil is designed to avoid small bending radii for
the HTS tape. The Lorentz force will push the compensation coil
outwards when the coil is powered. This movement will be supported by
expanding booms. In the expanded state, the compensation coil is in
stable mechanical equilibrium (Fig.~\ref{fig:magnet}). The very
small additional material budget of the compensation coil will have
negligible impact on the detector performance. Compensating the
magnetic dipole moment of the main solenoid requires a regulation of
the current in both magnets at the ppm level, similar to the precision
achieved for the current regulation of the LHC quadrupole
magnets~\cite{Thiesen2010}.

\section{AMS-100 Detector}
\subsection{Overview}
\begin{figure*}
  \centering
  \includegraphics[width=\textwidth]{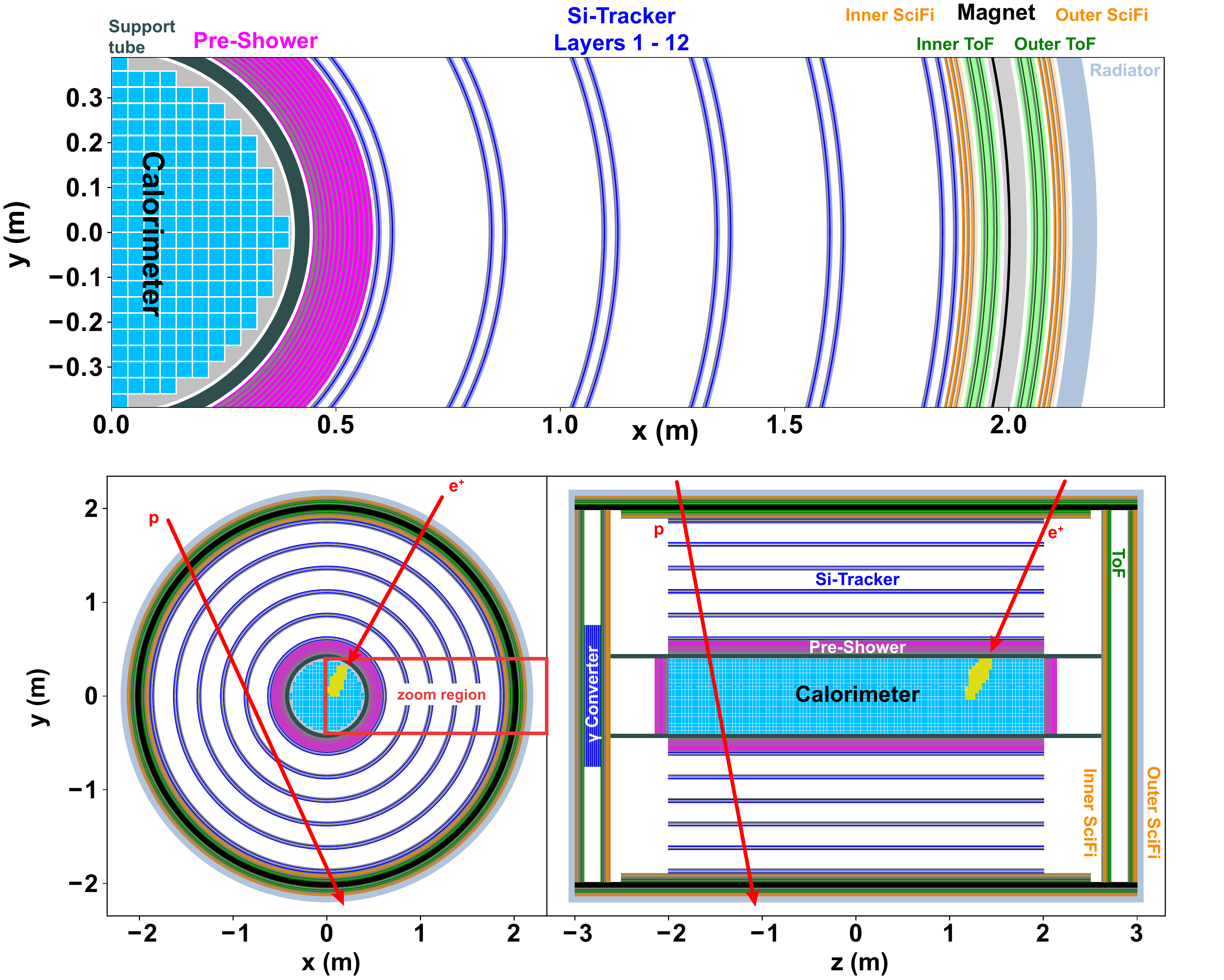}
  \caption[AMS-100 detector response.]{Schematic view of the AMS-100
    detector and its response to protons and positrons. The magnetic
    field inside the main solenoid is oriented in the $z$-direction,
    i.e.~the bottom left view shows the bending plane of the magnet,
    and a transverse view is shown on the bottom right. The upper
    panel shows a zoom into the bending plane view.
  \label{fig:response}}
\end{figure*}

The AMS-100 detector (Fig.~\ref{fig:response})
is located on the cold side behind the sunshield.

The main solenoid is instrumented both on the outside and on the
inside with a 3-layer high-resolution scintillating fiber (SciFi)
tracker~\cite{Beischer2010,Kirn2017} and a 2-layer time of flight
system (ToF). The SciFi tracker is assumed to have a
\hypertarget{q:scifi}{single point resolution} of \SI{40}{\um}. These
sub-detectors will provide fast information on the incoming particles,
as undistorted by the instrument as possible.

The inner detector consists of a silicon tracker, similar in design to
the AMS-02 silicon tracker~\cite{Alcaraz2008}, followed by a
pre-shower detector and a Lutetium-Yttrium oxyorthosilicate (LYSO)
crystal calorimeter~\cite{Zhang2014} with an outer radius of
\SI{40}{cm}. In addition to the SciFi-Tracker modules and
ToF-detectors, the endcap opposite the service module is instrumented
with photon converters to allow the reconstruction of low-energy
photons with good angular resolution. These converters consist of
silicon detector layers interleaved with thin tungsten layers as
proposed for GAMMA-400~\cite{Galper2018}.

AMS-100 has a \hypertarget{q:acceptance}{geometrical acceptance} of
\SI{100}{m^2.sr}, i.e.~1000 times the acceptance of AMS-02. The
instrument will monitor most of the sky continuously and will orbit
around the Sun in one year, together with Earth and L2. This will
guarantee homogenous sky coverage for $\gamma$-ray astronomy.  The
weight estimate of the instrument is given in
Table~\ref{tab:weight}. It has eight million
\hypertarget{q:chan}{readout channels} in total and an estimated total
\hypertarget{q:power}{power consumption} of \SI{15}{kW}.

\begin{table}
  \centering
  \begin{tabular}{lr}
 Component            & Weight (t) \\ \hline
 Tracking and ToF     &   5 \\
 Calorimeter          &  12 \\
 Main solenoid        &   1 \\
 Cabling              &   3 \\
 Cooling              &   3 \\
 Service module       &   2 \\
 Radiators            &   1 \\
 Sunshield            &   1 \\
 Support              &   9 \\
 Contingency          &   6 \\ \hline
 Total                &  43 \\
  \end{tabular}
  \caption{AMS-100 weight estimate.
  \label{tab:weight}}
\end{table}

\subsection{Event trigger}
Reducing the \SI{2}{MHz} \hypertarget{q:rawrate}{rate} of incoming
particles to an acceptable \hypertarget{q:trig}{level} of a few kHz
for the higher level DAQ systems and to a
\hypertarget{q:datarate}{data rate} of
$\sim$\SI{28}{Mbps}~\cite{sunshield} for the transfer to Earth with
on-board computers will be a major challenge. To overcome it, the fast
information provided by the outer detector (ToF-system and
SciFi-tracker) will be used for the trigger decisions, in combination
with calorimeter measurements: The track segments of the higher energy
particles reconstructed in the SciFi tracker will provide a first
estimate of the particle's rigidity up to the TV scale, and the ToF
signal amplitudes will determine the particle's charge. This will
allow the configuration of flexible trigger menus. For example, light
nuclei with rigidity below \SI{100}{GV} have to be mostly
rejected. Charged particles with an energy below $\sim\SI{100}{MeV}$
will be deflected by the magnetic field of the main solenoid and will
not be able to enter the detector volume. Prescaled random triggers
will be used to estimate the related trigger efficiencies. In
addition, those SciFi- and ToF-layers located outside the main
solenoid will be used to veto charged particles when reconstructing
$\gamma$ rays.

\subsection{Silicon tracker}
The silicon tracker is assumed to have a \hypertarget{q:si}{single
point resolution} of \SI{5}{\um} in the bending plane for $|Z|=1$
particles.  It consists of six double layers arranged in cylindrical
geometry (Fig.~\ref{fig:response}) leading to a maximum of 24
measurement points for a single track. For comparison, the CMS barrel
silicon tracker~\cite{CMS_Detector} has an outer radius of \SI{1.2}{m}
and consists of 10 layers, providing up to 20 measured points for a
cosmic muon going through the instrument. In combination with the
\SI{4}{m} diameter of the magnet and the magnetic field of \SI{1}{T},
the AMS-100 silicon tracker provides an MDR of \SI{100}{TV}.

\subsection{Time-of-Flight system}
To reconstruct particle masses and thus identify isotopes in cosmic
rays, a high-resolution ToF-system is required. Such systems constructed 
from small scintillator rods with time
resolution of \SIrange{30}{50}{ps} are presently under
construction~\cite{cmstof,pandatof}. We assume here that the time
resolution of the PANDA ToF can be significantly improved using a
larger coverage of the scintillator rods with SiPMs and operating the
detector at \SI{200}{\kelvin}. For $|Z|=1$ particles, we target for a
time resolution of \SI{20}{ps} for a single scintillator rod leading
to a \hypertarget{q:tof}{time resolution} of \SI{15}{ps} for the
4-layer ToF system.

\subsection{Calorimetry}
\begin{figure}
  \centering
  \includegraphics[width=\columnwidth]{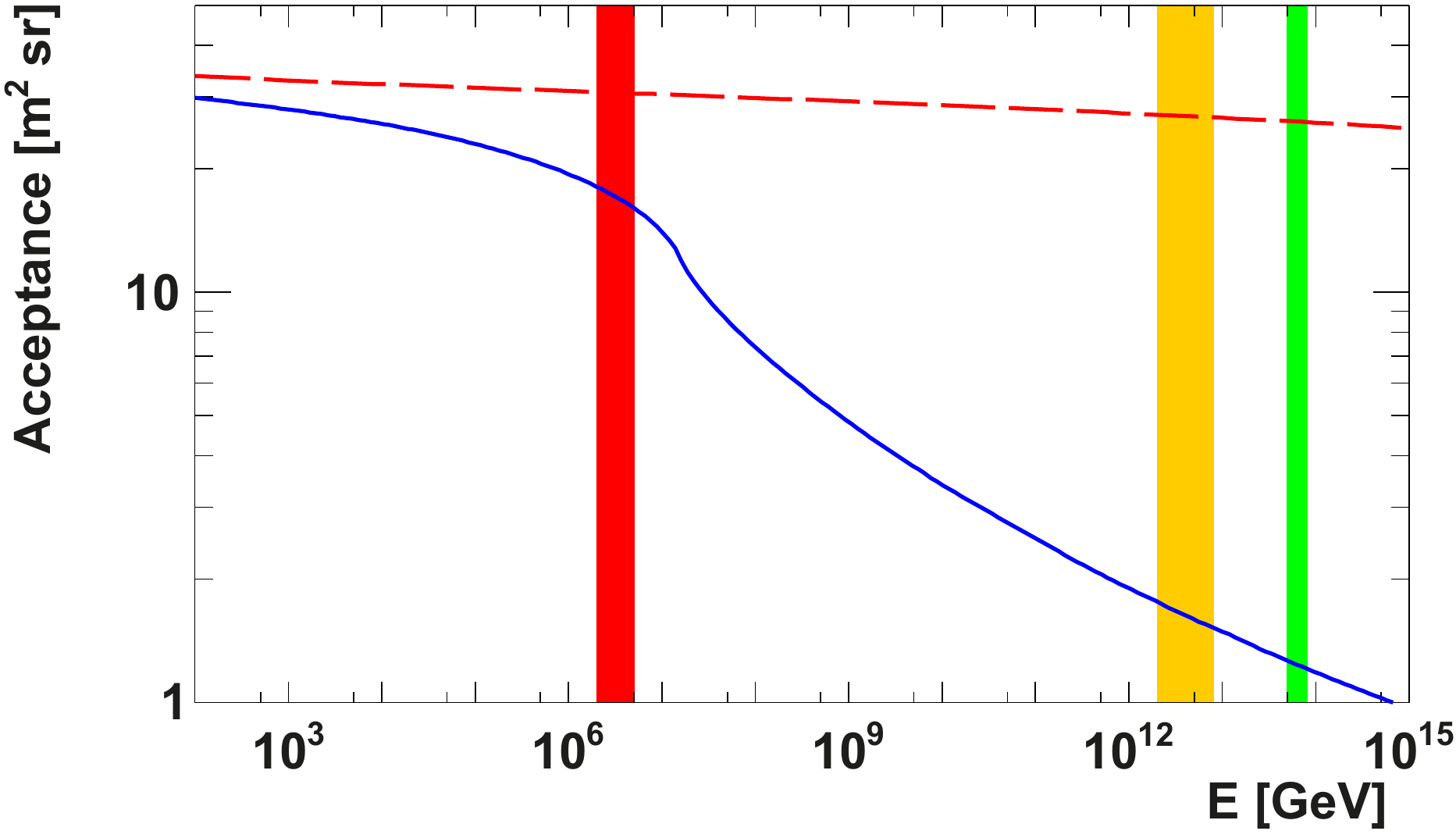}
  \caption[Acceptance of calorimeter system.]{Acceptance of the
    calorimeter system for hadronic showers (solid blue curve) and
    electromagnetic showers (dashed red curve) as a function of
    energy. We assume here that for a useful measurement the maximum
    of the shower needs to be contained in the calorimeter, which
    has a maximum depth of $20\,\lambda_I$  
    along the $z$-axis, and $4\,\lambda_I$ along the diameter.
    The effective thickness depends on the track angle and impact point 
    at the outer radius of the calorimeter.
    The red band indicates the energy of
    the cosmic-ray knee, the yellow one the energy of the ankle and
    the green one the GZK cutoff energy.}
  \label{fig:acccalo}
\end{figure}

The pre-shower detector and the LYSO crystal calorimeter are used to
separate electromagnetic and hadronic showers, and to measure the
energy of electrons, positrons and photons, as well as protons and
ions beyond the MDR. The crystal calorimeter is inspired by the design
of the HERD detector~\cite{Zhang2014} and allows the three-dimensional
reconstruction of the shower shape. The pre-shower detector consists
of 12 silicon detector layers interleaved with thin tungsten layers to
provide good angular resolution for the measurement of $\gamma$ rays
and to limit the backsplash of the calorimeter into the silicon
tracker. This combination of pre-shower detector and crystal
calorimeter has a \hypertarget{q:calo}{depth} of $70\,X_0$, or
$4\,\lambda_I$, for particles incident in the bending plane of the main
solenoid and hitting the calorimeter centrally. The geometrical
acceptance of this system allows the measurement of cosmic nuclei with
energies above \SI{100}{TV} up to the cosmic-ray knee at the PeV scale
(Fig.~\ref{fig:acccalo}). With today's accelerators, AMS-100 can only
be calibrated up to \SI{400}{GeV}. In orbit, the energy scale of the
calorimeter system will be calibrated in the energy range from
\SI{100}{GeV} to \SI{100}{TeV} using the rigidity measurement of
charged cosmic rays in the spectrometer.

\subsection{Support tube and service module}
The main structural element is a central \SI{3}{cm} thick carbon
support tube with an outer radius of \SI{44}{cm} around the
calorimeter. It will mechanically stabilize the detector during the
launch and connect the service module to the launch adapter, which is
the interface to the rocket. The main solenoid and the other
subdetectors are connected to the central support tube by lightweight
carbon fiber structures. Services are routed in the volumes between
the barrel and the endcap detectors to the service module. The service
module accommodates the DAQ system, the power distribution system, the
telecommunication system, the attitude control, the thermal control
system, and an electric propulsion system to keep a stable orbit
around L2. A combination of reaction wheels and electric propulsion is
used to keep the orientation of the sunshield stable with respect to
the Sun.

\subsection{Sunshield}
The sunshield has a radius of \SI{9}{m} and is designed similar to the
concept developed for the James Webb Space
Telescope~\cite{sunshield}. The dimensions of the sunshield are chosen
such that a pointing accuracy of a few degrees towards the Sun is
sufficient to keep the magnet system cool. Other than for thermal
reasons, the orientation of the instrument has no impact on the
physics program. Star trackers will be used to monitor the orientation
to provide precision information for the $\gamma$-ray astronomy
program.

\section{AMS-100 Physics Program}
This paragraph can only cover first ideas related to the AMS-100
physics program, a lot of new aspects have to be worked out in more
detail. This includes the sensitivity to various isotopes in cosmic
rays, heavy nuclei beyond iron in cosmic rays,
strangelets~\cite{doi:10.1142/S0217751X05029939}, magnetic
monopoles~\cite{PhysRevD.98.030001}, particles with fractional
charges~\cite{FUKE20082050}, evaporating primordial black
holes~\cite{hawking1975,PhysRevLett.76.3474}, search for signatures of
dark matter annihilation or decay in $\gamma$-ray
lines~\cite{Bergstrom:1997fj,Bergstrom:2012vd}, search for
axions~\cite{Raffelt:1987im,DeAngelis:2007dqd}, or tests of quantum
gravity by precisely measuring the energy and arrival time of photons
from $\gamma$-ray bursts~\cite{AmelinoCamelia:1998th}, to mention a
few examples that can be covered in principle with unprecedented
sensitivity by such a powerful instrument.

For the following performance estimates, the detector acceptances have
been determined with the help of a
\texttt{Geant4}~\cite{Agostinelli2003} simulation.

\subsection{Protons and heavier nuclei}
\begin{figure}
  \centering
  \includegraphics[width=\columnwidth]{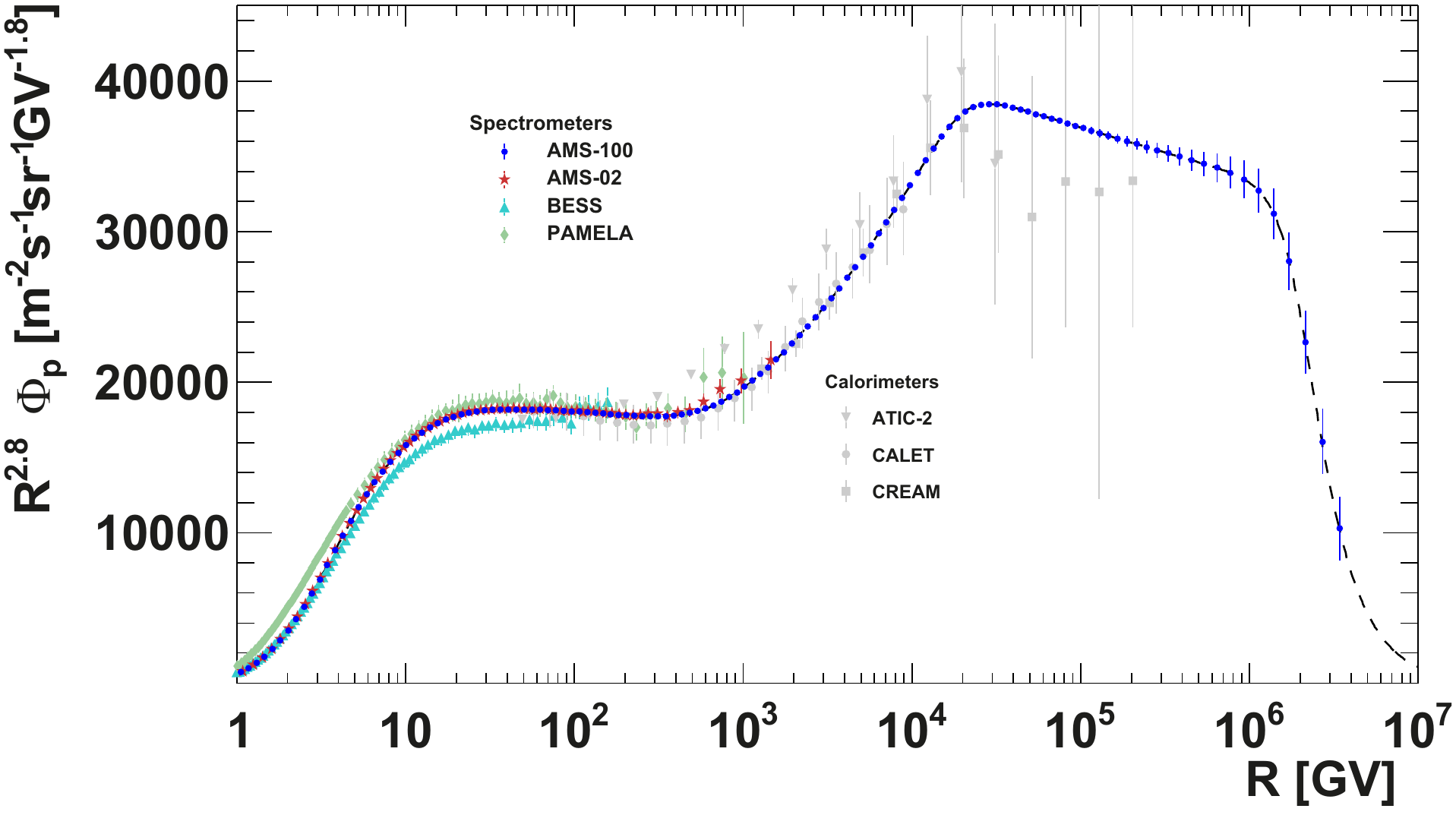}
  \caption[Cosmic-ray proton spectrum.]{Cosmic-ray proton
    spectrum. Expected data from AMS-100 (blue) (statistical
    uncertainties only), for the case that the proton flux is
    described by a power law with several smooth breaks,
    inserted for the purpose of illustration (dashed curve).
    Recent magnetic spectrometer measurements from
    BESS~\cite{Abe_2016}, PAMELA~\cite{Adriani:2011hja}, and
    AMS-02~\cite{Aguilar2015}. Recent calorimeter measurements from
    ATIC-2~\cite{Panov2009}, CALET~\cite{PhysRevLett.122.181102}, and
    CREAM-III~\cite{Yoon2017}.}
  \label{fig:protons}
\end{figure}

Protons are the most abundant species in cosmic rays. PAMELA and
AMS-02 have reported a spectral break above $\sim$\SI{200}{GV} in
protons and other light
nuclei~\cite{Aguilar2015,Adriani2011b,Aguilar2017}. Spectral breaks
encode information about the sources and the propagation of cosmic
rays~\cite{Evoli:2018nmb,Genolini:2017dfb}. Up to now there is no
coherent description of the various features observed in the energy
spectra of cosmic rays. AMS-100 will measure protons and heavier
nuclei in cosmic rays up to the maximum energy that can be reached by
galactic cosmic-ray accelerators (Fig.~\ref{fig:protons}). The
positions of the spectral features in the spectra of different
species, as well as the dependence of their appearance on the nucleus
charge should provide the most detailed information about the
cosmic-ray sources and processes in the interstellar medium. This
information forms the necessary basis for other studies detailed
below, such as the origin of cosmic-ray positrons, electrons,
antiprotons, and antimatter. In addition, these direct measurements at
the highest energies will allow us to investigate the change of the
chemical composition of cosmic rays at the knee and gather invaluable
information about the transition from Galactic to extragalactic cosmic
rays.

\subsection{Positrons and Electrons}
The observed excess of high-energy positrons above the
expected yield from cosmic-ray collisions has generated widespread
interest and discussions. Possible interpretations range from new
effects in the acceleration and propagation of cosmic
rays~\cite{Lipari:2016vqk,Cowsik:2013woa,Blum:2013zsa} to acceleration
of positrons to high energies in astrophysical
objects~\cite{Fujita:2009wk,Serpico:2011wg,Linden:2013mqa,Mertsch:2014kd,Tomassetti:2015cva,Hooper:2017gtd,Liu:2016gyv,Kachelriess:2017yzq,Profumo:2018fmz}
and to dark
matter~\cite{Turner:1989kg,Ellis:1999yw,Cheng:2002ej,Cirelli:2008pk,Kane:2009if,Kopp:2013eka,Chen:2015cqa,Cheng:2016slx,Bai:2017fav}
as a new source of cosmic-ray positrons. The latest data on the
positron flux from AMS-02 show a spectral break at \SI{300}{GeV}
followed by a sharp drop~\cite{Aguilar2019}. The detailed
understanding of the shape of the spectrum above this energy is the
key to deduce the origin of these high energy positrons.

\begin{figure}
  \centering
  \includegraphics[width=\columnwidth]{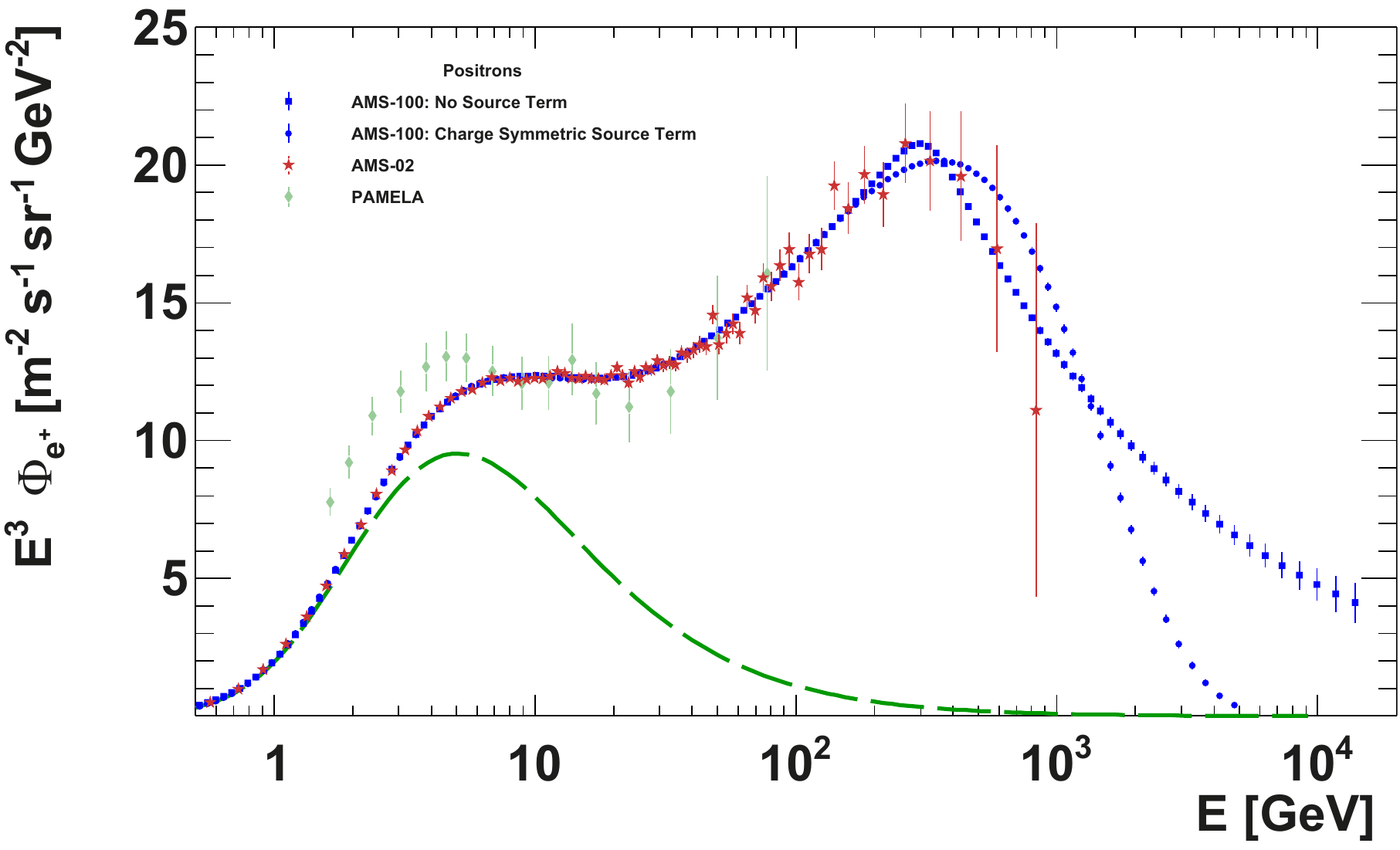}
  \caption[Cosmic-ray positron spectrum.]{Cosmic-ray positron
    spectrum. Expected data from AMS-100
    (stat.~uncertainties only) for two different scenarios: a) The
    spectrum is described by a power law plus a source term with an
    exponential cutoff (blue circles, lower curve at high energy).
    b) The spectrum is
    described by power laws with spectral breaks and the last break is
    at \SI{300}{GeV} (blue squares, upper curve at high energy).
    The dashed green curve shows the
    expected spectrum from a) without the source term. 
    Recent experimental data from PAMELA~\cite{PhysRevLett.111.081102} and 
    AMS-02~\cite{Aguilar2019} are shown.
    \label{fig:positrons}}
\end{figure}

A generic source term, that describes the contribution of the new
source responsible for the positron excess, is given by a power law
with an exponential cutoff (e.g., Ref.~\cite{Aguilar2019}). AMS-100
will be able to precisely measure the cosmic-ray positron spectrum up
to \SI{10}{TeV} (Fig.~\ref{fig:positrons}).

\begin{figure}
  \centering
  \includegraphics[width=\columnwidth]{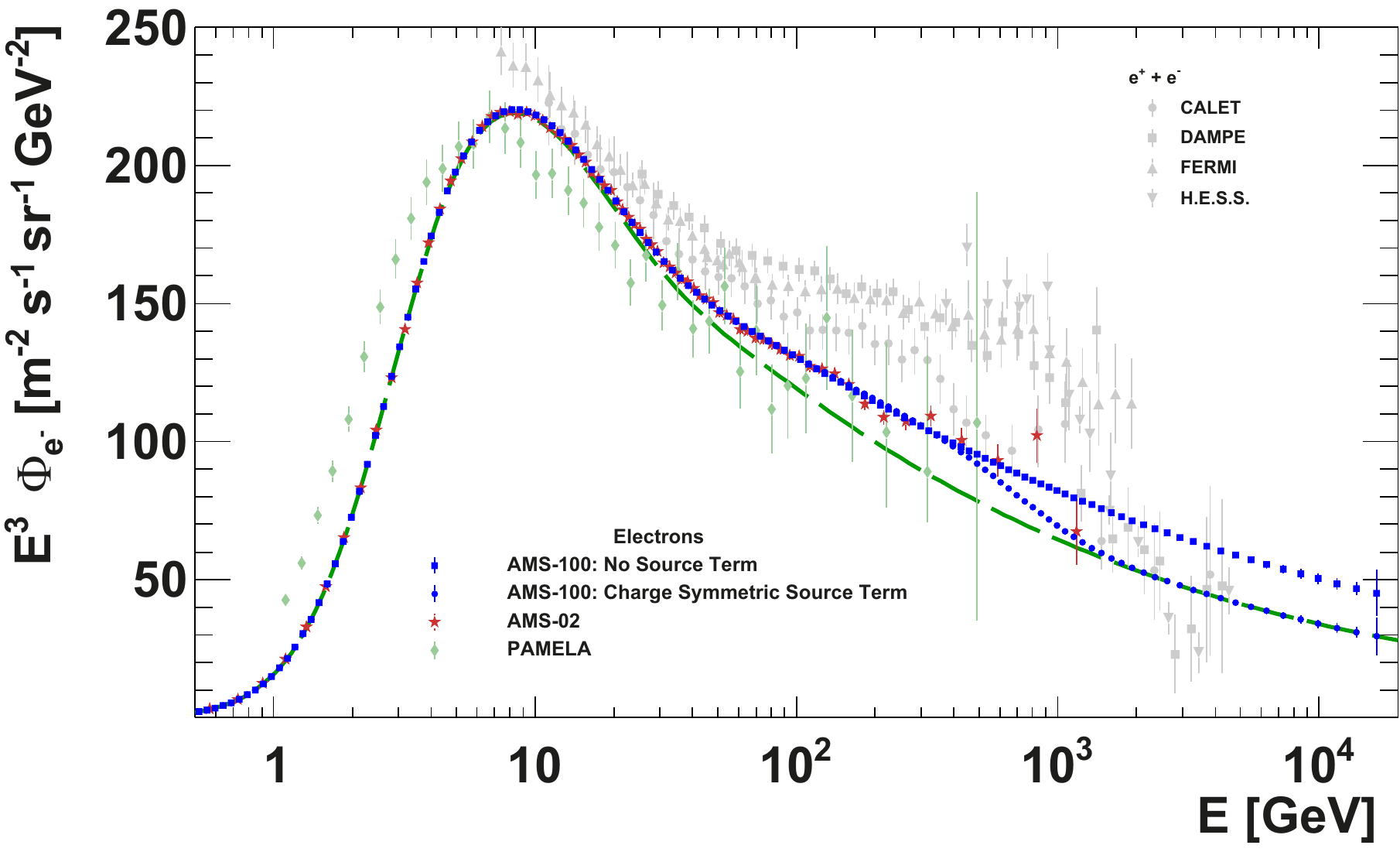}
  \caption[Cosmic-ray electron spectrum.]{Cosmic-ray electron
    spectrum. Expected data
    from AMS-100 in blue (stat.~uncertainties only) for two different
    scenarios: a) Broken power law plus a charge symmetric source term
    as obtained a fit to the positron flux (blue circles, lower curve
    at high energy). b) The
    broken power law continues without any further spectral break at
    high energies (blue squares, upper curve at high energy).
    The dashed green curve shows the
    derived spectrum from a) without the source term. 
    Recent experimental data from PAMELA~\cite{PhysRevLett.106.201101} and 
    AMS-02~\cite{Aguilar2019a} are shown. For comparison
    also the recent calorimetric measurements of the combined $(e^+ +
    e^-)$ flux by CALET~\cite{PhysRevLett.120.261102}, 
    DAMPE~\cite{Ambrosi2017}, FERMI~\cite{Abdollahi:2017ki}, 
    and H.E.S.S.~\cite{Aharonian2009b,PhysRevLett.101.261104}
    are shown as they extend 
    to higher energies and provide an upper limit for the electron flux. 
    \label{fig:electrons}}
\end{figure}

If the origin of the source term is a process producing electrons and
positrons in equal amounts, the effect should also be detectable in
the cosmic-ray electron spectrum. Both pulsar models and dark matter
models generically predict such a charge-symmetric source term.
H.E.S.S.~\cite{Aharonian2009b} and DAMPE~\cite{Ambrosi2017} have
observed a spectral break of the combined electron and positron flux
at about \SI{1}{TeV} followed by a sharp drop, which might be related
to this question. AMS-100 will be able to precisely measure the
cosmic-ray electron spectrum up to \SI{20}{TeV}
(Fig.~\ref{fig:electrons}) and detect features associated with the
local sources of electrons predicted in propagation
models. Identifying such features will shed light on the origin of
positrons, electrons, and other cosmic-ray species.

\subsection{Antiprotons}
\begin{figure}
  \centering
  \includegraphics[width=\columnwidth]{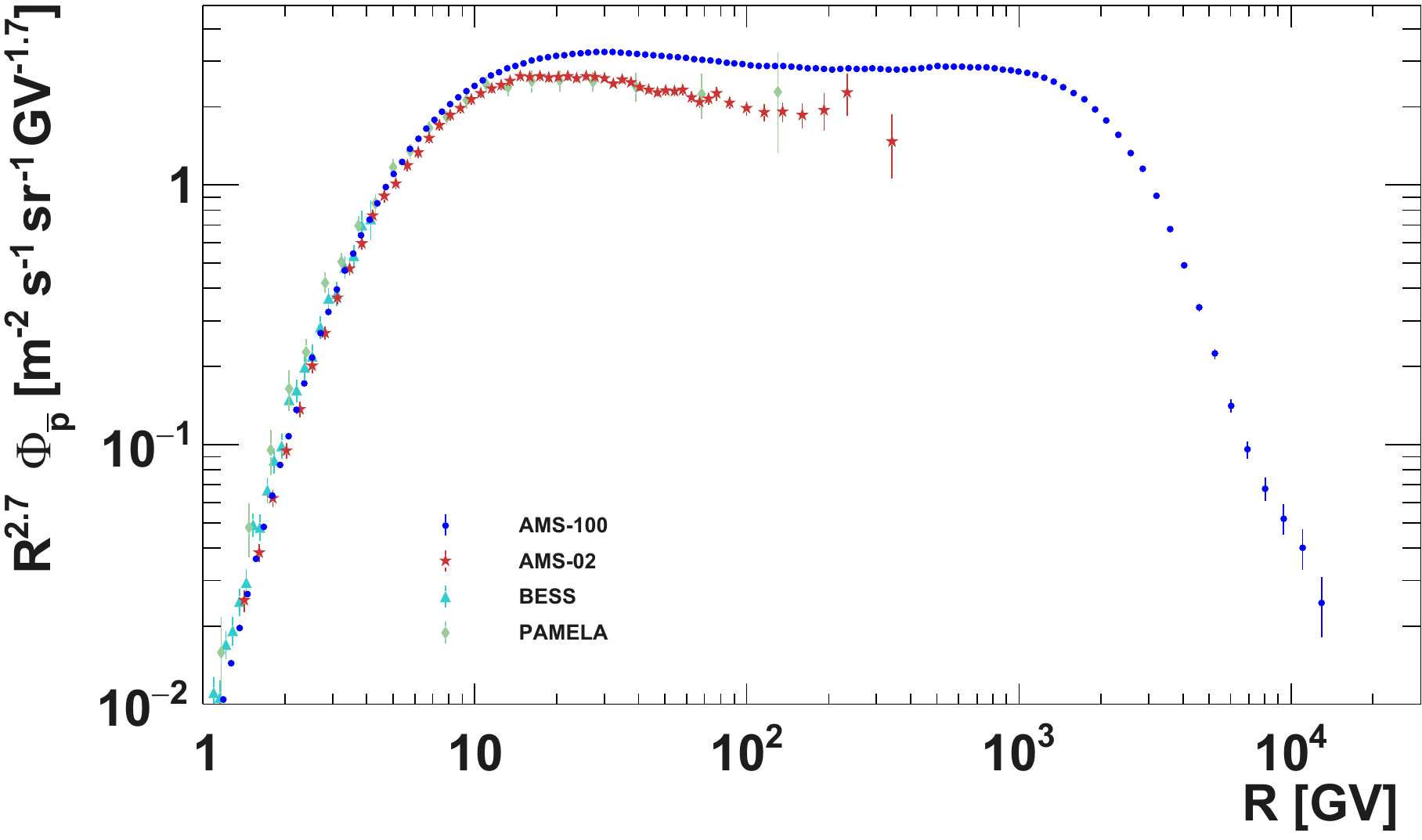}
  \caption[Cosmic-ray antiproton spectrum.]{Cosmic-ray antiproton
    spectrum. Recent experimental data from BESS-Polar~\cite{Abe2012a}, 
    PAMELA~\cite{Adriani2013} and AMS-02~\cite{Aguilar2016},
    together with the expected data from AMS-100 (blue)
    (stat.~uncertainties only) based on a model 
    prediction~\cite{Mertsch:2014kd}
    which was published before the AMS-02 data.
  \label{fig:pbar}}
\end{figure}

Positrons and electrons could be generated by a pulsar, but not
antiprotons. Antiprotons can only be produced in high-energy
interactions or in the annihilation of dark matter
particles. Therefore, antiproton measurements may provide support to
the dark matter hypothesis for the origin of the positron excess or
rule it out. Independently, they provide another crucial probe of the
processes in the interstellar medium, as well as production and
acceleration of secondary species in the
sources~\cite{Johannesson:2016rlh}. AMS-100 will be able to measure
the antiproton spectrum up to the \SI{10}{TeV} energy scale and
provide precise information on the spectral shape. Hence it will shed
light on many questions associated with the origin of cosmic rays and
with the nature of dark matter (Fig.~\ref{fig:pbar}).

\subsection{Antihelium}
AMS-02 has shown both $\hebart$ and $\hebarf$ candidate events at
conferences~\cite{hebar}. These unexpected events are observed in
AMS-02 at a rate of 1~event/year or 1~$\hebar$ event in 100~million He
events. The rate of secondary $\hebar$ nuclei predicted by coalescence
models is significantly lower. Therefore, the origin of the $\hebar$
nuclei is unclear. The independent confirmation of these candidate
events would have the most profound implications for physics and
astrophysics. Besides the question of the statistical significance of
the signal, the independent systematic uncertainties of the new
instrument are essential. This requires an instrument with a different
detector design at a different location in space. Extrapolating the
AMS-02 $\hebar$ event rate to the AMS-100 acceptance results in the
prediction of finding in the order of 1000 $\hebar$ events/year. The
precision measurement of the spectral shape of the $\hebar$ flux would
allow tests of the origin of $\hebar$. The rotational symmetry of
AMS-100 allows detailed systematic cross-checks of such a result
equivalent to inverting the magnetic field.

\subsection{Antideuterons}
\begin{figure}
  \centering
  \includegraphics[width=\columnwidth]{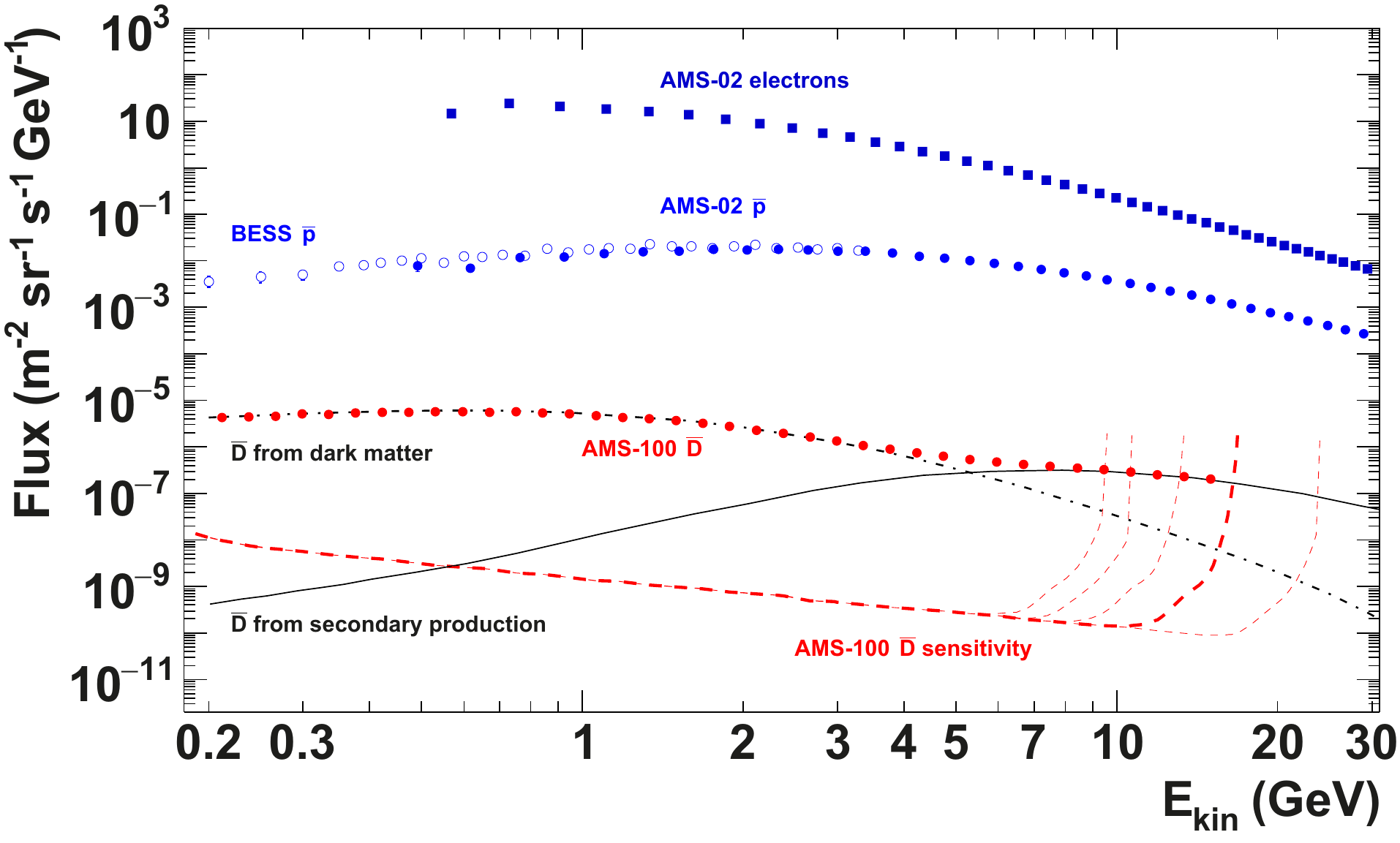}
  \caption[Sensitivity to antideuterons.]{Differential sensitivity of
    AMS-100 to antideuterons in cosmic rays for a mission time of 10
    years and a single-layer ToF time resolution of \SI{20}{ps}, with
    a logarithmic binning of 20 bins per decade (thick
    dashed red curve). Sensitivities for time resolutions of
    \SIlist{10;30;40;50}{ps} are shown by thin dashed red curves (from
    right to left). The red
    symbols show the expected data from AMS-100 if the antideuteron
    flux follows the dark matter model of Ref.~\cite{Korsmeier2018}
    with statistical uncertainties (which are smaller than the symbol
    size).
    The solid black curve shows the
    antideuteron flux expected from secondary production by charged
    cosmic rays interacting with the interstellar material, as derived
    in Ref.~\cite{Lin2018} for the EPOS LHC interaction model. Data
    for the other $Z=-1$ particles in cosmic rays, from
    AMS-02~\cite{Aguilar2016,Aguilar2014} and
    BESS-Polar~\cite{Abe2012a}, are shown to indicate the signal to
    background ratios for the antideuteron measurement.
    \label{fig:dbar}}
\end{figure}

\begin{table}
  \centering
  \begin{tabular}{lccc}
    Experiment              & Energy range  & $\bar{D}$ sensitivity   & Ref. \\ 
                            & (\si{GeV/n})  & (\si{[m^{2}.s.sr.GeV/n]^{-1}})  & \\ \hline
    GAPS                    & $0.1$ to $0.25$  & $2.0\cdot10^{-6}$   & \cite{Aramaki2016} \\ \hline
    \multirow{2}{*}{AMS-02} & $0.2$ to $0.8$   & $4.5\cdot10^{-7}$   & \cite{Choutko2008} \\
                            & $2.2$ to $4.2$   & $4.5\cdot10^{-7}$   & \cite{Choutko2008} \\ \hline
    AMS-100                 & $0.1$ to $8.0$   & $3\cdot10^{-11}$  & \\
  \end{tabular}
  \caption[Comparison of antideuteron sensitivities.]{Comparison of
    antideuteron sensitivities. (The AMS-02 sensitivity was estimated
    in Ref.~\cite{Choutko2008} for the superconducting magnet instead
    of the permanent magnet used in the flight configuration.)}
  \label{tab:antid}
\end{table}
Antideuterons potentially are the most sensitive probe for dark matter
in cosmic rays~\cite{Donato:1999gy,Cui2010}. While antiprotons are
predominantly produced in secondary interactions in the interstellar
medium, antideuterons at low energy have no other known origin. No
antideuterons have ever been identified in cosmic rays. The current
best limit has been set by BESS~\cite{Fuke2005}, excluding a flux of
\SI{1.9e-4}{(m^{2}.s.sr.GeV/n)^{-1}} between \SI{0.17}{GeV/n} and
\SI{1.15}{GeV/n} at the \SI{95}{\percent} confidence level. The
expected sensitivity of AMS-100 is \num{3e-11}
\si{(m^{2}.s.sr.GeV/n)^{-1}} in the energy range between
\SI{0.1}{GeV/n} and \SI{8}{GeV/n}. It is compared to other experiments
in Table~\ref{tab:antid}. At this level of sensitivity, it is no
longer useful to quote an {\em integral} sensitivity, which is related
to the chances of observing a certain number of events {\em anywhere}
inside a given energy range. Instead, we calculate a {\em
  differential} sensitivity, which can be directly compared to model
predictions for the differential $\bar{D}$ flux. We choose a
logarithmic energy binning with 20 bins per decade and calculate the
sensitivity individually for each bin. It is defined as the
\SI{95}{\percent} confidence level limit that will be set in case no
$\bar{D}$ events are observed in the given bin. The differential
sensitivity for antideuterons is shown in Fig.~\ref{fig:dbar}. AMS-100
will be the first instrument to measure the cosmic-ray antideuteron
spectrum with thousands of events, even in the case that antideuterons
originate only from secondary production. AMS-100 will have the
sensitivity to distinguish between antideuterons originating in dark
matter annihilations and those produced in interactions within the
interstellar medium, due to the different spectral shapes expected for
these components. While it is not clear if antideuterons from dark
matter annihilation exist, the observation of antideuterons from
secondary production would allow us to set additional constraints on
the $\hebart$ and $\hebarf$ rates in cosmic rays: Within the
coalescence model~\cite{Chardonnet1997}, every nucleon in the
antimatter particle reduces the production rate by a factor
$\simeq{}10^3\mhyphen{}10^4$ depending on the energy, i.e.~we expect
$N(\bar{p}):N(\bar{D}):N(\hebart):N(\hebarf)\approx1:10^{-3}\mhyphen10^{-4}:10^{-6}\mhyphen10^{-7}:10^{-9}\mhyphen10^{-10}$
in cosmic rays if there is no new source for one of these antimatter
species. A simultaneous measurement of these sensitive probes for new
physics is therefore required to derive a coherent picture.

\subsection{High-energy Gamma Rays}
Building on the success of current-generation $\gamma$-ray detectors
such as Fermi-LAT~\cite{3fhl}, AMS-100 will allow detailed studies of
$\gamma$-ray sources and the diffuse $\gamma$-ray emission up to the
$\simeq$ 10~TeV scale. It has an acceptance of \SI{30}{m^2.sr} for photons
reconstructed in the calorimeter system. Due to the pre-shower
detector, the expected angular resolution is compatible to the one of
Fermi-LAT. In addition, a similar acceptance is obtained from photon
conversions in the thin main solenoid, resulting in a total acceptance
for diffuse photons of up to \SI{60}{m^2.sr}.

\begin{figure*}
  \centering
  \includegraphics[width=0.51\textwidth]{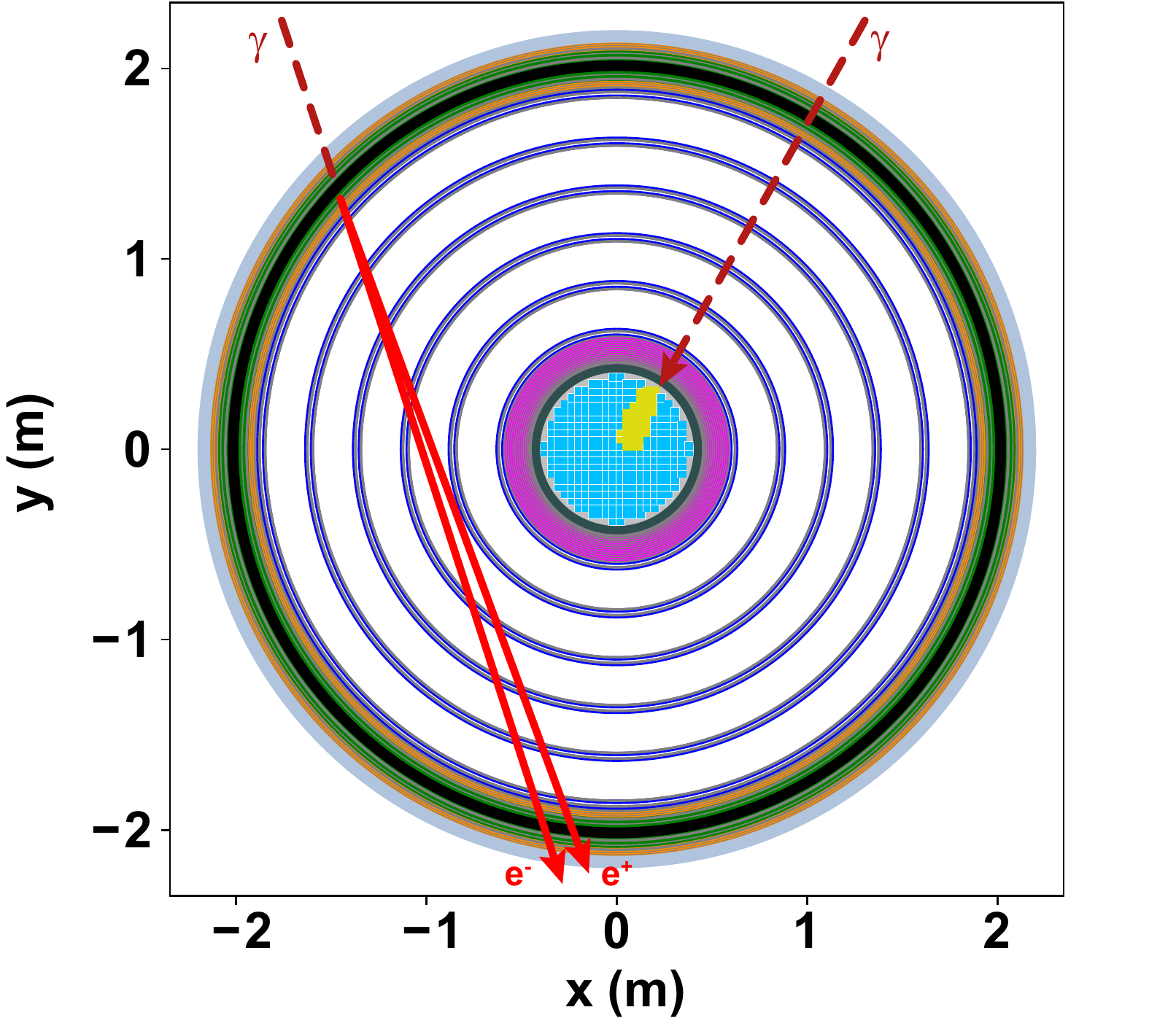}
  \includegraphics[width=0.48\textwidth]{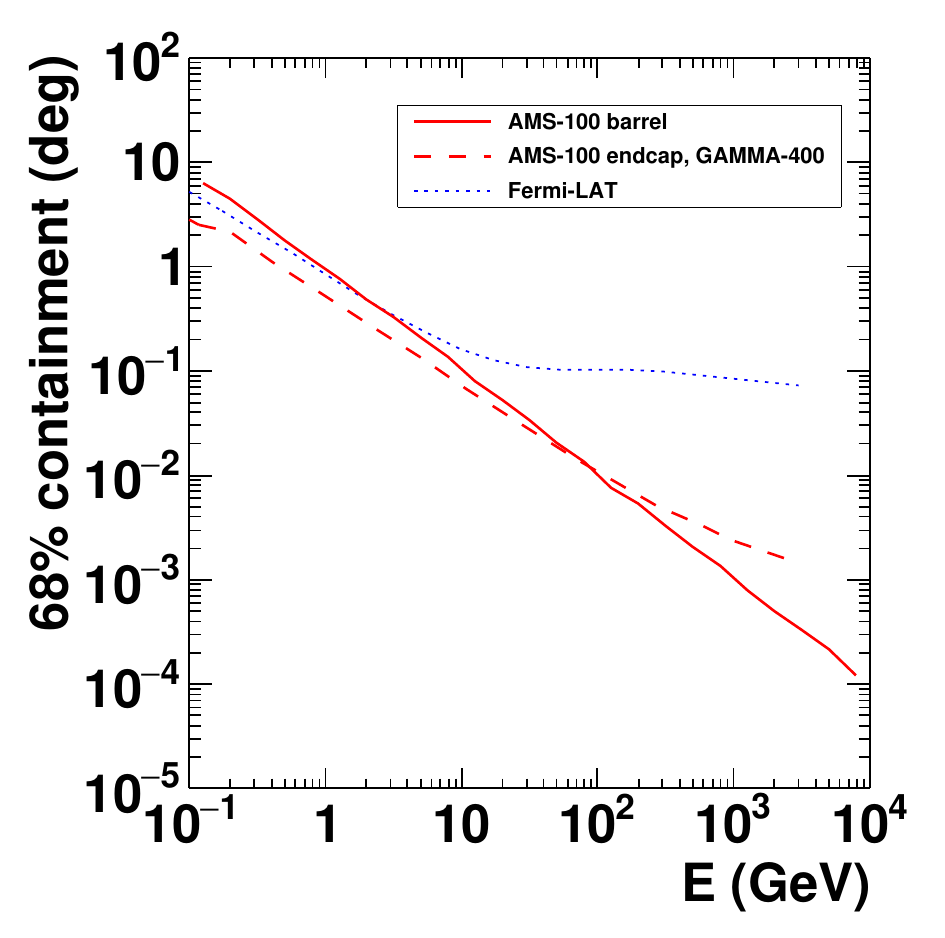}
  \caption[Expected angular resolution for photons.]{
    Left: Sketch of a $\gamma$ conversion in the AMS-100 main
    solenoid and of a $\gamma$ reconstructed in the calorimeter.
    Right: Expected angular resolution (68\% containment)
    for photons converted in the AMS-100 barrel, based on a
    \texttt{Geant4} simulation of the 
    multiple scattering in the detector material, and for the endcap
    detector which follows the design of GAMMA-400~\cite{Galper2017}.
    The resolution function of Fermi-LAT~\cite{fermi_angres} is shown
    for comparison.
    \label{fig:photons}}
\end{figure*}

At low energies the angular resolution for converted photons is
limited by multiple scattering of the resulting electron-positron
pairs. But at high energies, the direction of the photon can be
reconstructed with high accuracy due to the good spatial resolution
and long lever arm of the silicon tracker (Fig.~\ref{fig:photons}).
This will resolve structures in $\gamma$-ray sources with angular
resolution similar to today's best X-ray telescopes. Interesting
targets include galactic supernova
remnants~\cite{Aharonian2013,Funk2017}, pulsar wind
nebulae~\cite{Gaensler2006}, and blazars. For converted photons
perpendicular to the $z$-axis the effective area reaches
\SI{2.5}{m^2}.

Due to the rotational symmetry of its barrel detector, its dedicated
endcap photon detector, and its location far from the shadow of the
Earth, AMS-100 will be able to monitor almost the entire sky
continuously. Combined with its large effective area, this will make
it a prime instrument for instantaneous observation of transient
sources, e.g.~$\gamma$-ray bursts or photons emitted in conjunction
with gravitational wave events, as well as for monitoring blazar
variability~\cite{blazars}. In combination with ground-based
experiments, it will allow completing the multi-messenger network for
modern astronomy combining the observation of gravitational waves,
cosmic-ray neutrinos and GeV-TeV $\gamma$ rays. AMS-100 can serve as a
trigger for the Cherenkov Telescope Array~\cite{Acharya2013} and
similar ground-based observatories for the detailed follow-up
investigation of transient sources.

The physics program of AMS-100 covering galactic and extragalactic
$\gamma$-ray sources will be detailed in future publications. One
example is the study of $\gamma$-ray pair halos around blazars,
e.g.~\cite{Neronov:1900zz}. TeV $\gamma$ rays emitted from the jets of
blazars produce pairs of electrons and positrons through interactions
with the extragalactic background light (EBL). These electrons and
positrons further lose their energy through synchrotron and inverse
Compton emission, thus initiating a cascade of lower-energy electrons,
positrons and $\gamma$ rays. Depending on the properties of the
intergalactic magnetic fields, $\gamma$ rays from such cascades
can be observed in the form of extended $\gamma$-ray halos. With its
improved diffuse sensitivity, AMS-100 would be able to detect or
constrain the existence of pair halos and thus put new bounds on the
strength and correlation length of the intergalactic magnetic field.

One can also search for spectral features in the $\gamma$-ray emission
of blazars due to attenuation from the EBL. This allows drawing
conclusions on axion-photon couplings~\cite{Raffelt:1987im,DeAngelis:2007dqd}. Measuring blazar
spectra to higher energies with AMS-100 extends the sensitive
parameter space to smaller couplings.\newline
\par
The excellent timing and pointing capabilities of AMS-100 make it an
ideal instrument to test Lorentz invariance violation (LIV) by
precisely measuring the energy and arrival time of photons from
$\gamma$-ray bursts~\cite{AmelinoCamelia:1998th}. Deviations of the
group velocity of photons from the speed of light, which could be
realised in models of quantum gravity, would manifest themselves in
different arrival times of photons of different energies from the same
transient event. Given the energy reach of AMS-100, the
observation of very high-energy $\gamma$ rays in conjunction with X-ray
instruments would increase the sensitivity to LIV by
orders of magnitude compared to existing measurements.

\subsection{AMS-100 Pathfinder}
The technical complexity of the AMS-100 project requires a pathfinder
mission, similar to the AMS-01 flight on Space Shuttle Discovery in
1998~\cite{AMS-01}, or to the ongoing LISA program. This pathfinder
mission has to demonstrate the stable operation of a HTS magnet in
space for the first time, including the expandable compensation coil
technology.  It has to be operated at L2 to verify the
thermo-mechanical design and to demonstrate the sufficient attitude
control inside the time-varying interplanetary magnetic field. Testing
the quench probability of the magnet system in this environment and
the impact of a quench on the instrument is of key importance. The
successful test will qualify similar HTS magnet configurations as
radiation shield for a crew compartment for interplanetary manned
space flights as discussed in Ref.~\cite{NASA-HTS}.

Given the effort of a space mission at L2, a purely technical
demonstrator mission would be a waste of resources. Therefore the
AMS-100 pathfinder is anticipated to be a prototype at the 10\% scale
level of AMS-100, i.e. the length and the radius of the main solenoid
are reduced by a factor 2 to $L=\SI{3}{m}$ and $R=\SI{2}{m}$, so that
the instrumented volume is reduced by nearly an order of
magnitude. Its weight is estimated to be \SI{5}{t} and its detector
concept is in all other aspects very similar to AMS-100. The central
calorimeter has to be removed due to weight constraints as other
components like the service module do not scale accordingly. With
these dimensions and weight, the AMS-100 pathfinder can be launched to
L2 with an Ariane~5 or a rocket of similar scale.

For the physics program of the pathfinder mission, the key performance
parameters are a geometrical acceptance of \SI{20}{m^2 sr} and an MDR
of \SI{5}{TV}. The sensitivity for heavy cosmic antimatter particles
would be reduced compared to AMS-100 by an order of magnitude, but
compared to AMS-02 this 10\% scale pathfinder already has a
$100\times$ higher sensitivity to heavy cosmic antimatter particles
and completely independent systematic uncertainties, due to its
different detector geometry, detector technology and orbit.

\subsection{Cost estimates and timeline}
\begin{table}
  \centering
  \begin{tabular}{lc}
    R\&D phase                              &  2019 - 2021\\
    Detailed technical design report        &  2020 - 2022\\
    Construction phase AMS-100 Pathfinder   &  2023 - 2028\\
    Launch AMS-100 Pathfinder               &  2029\\
    Science AMS-100 Pathfinder              &  2030 - 2036\\
    Construction phase AMS-100              &  2031 - 2038\\
    Launch AMS-100                          &  2039\\
    Science AMS-100                         &  2040 - 2050\\
  \end{tabular}
  \caption[AMS-100 Planning.]{Estimated schedule for the AMS-100 project.}
  \label{tab:schedule}
\end{table}

The AMS-100 project falls into the ESA or NASA class L category,
i.e.~the full mission requires a budget of more than 1 billion dollars.
The scale of the project requires a large international collaboration
as successfully demonstrated by the AMS-02 project on the
International Space Station. The AMS-100 pathfinder mission falls
into the ESA class M category or NASA class S category, i.e.~it
requires a budget below 500 million dollars, with an estimated
instrument cost of 150 million dollars.

A possible timeline for the AMS-100 project is given in
table~\ref{tab:schedule}. The important milestones for the R\&D-Phase
are the first successful space qualification test of a high
temperature superconducting solenoid and the verification of the
achievable time resolution of the ToF system. The detailed
technical design report requires a valid thermo-mechanical model for
the mission including a detailed concept of the detector electronics,
DAQ system and data handling.

We welcome and invite contributions from interested groups with the
goal of participating in the R\&D-Phase and creating the technical
design report for the AMS-100 project.

\section{Summary}
The only magnetic spectrometer in space today, AMS-02, has collected
more than 140 billion cosmic rays since 2011 and will continue to take
data for the lifetime of the ISS, i.e.~the next decade. AMS-100 is an
ambitious project for the following decade which requires pushing
today's technology to its limits in several fields. Many demanding
technical questions need to be worked out in detail to make such a
large space mission possible. These questions are of similar
complexity as the ones that had to be solved to realize AMS-02 after
the proposal in 1994~\cite{Ahlen1994}.  The AMS-100 concept as
outlined in this article (Tab.~\ref{tab:quantities}) has the potential
to improve the sensitivity of AMS-02 by a factor of 1000. This means
that we will reproduce 20 years of AMS-02 data within the first week
of operation at Lagrange Point 2. In the second week, we will start
exploring completely new territory in precision cosmic-ray physics.

\begin{table*}
  \centering
  \begin{tabular}{lrl}
    Quantity                                   & Value                                      &                                                  \\ \hline
    \hyperlink{q:acceptance}{Acceptance}       & \SI{100}{m^2.sr}                           &                                                  \\
    \hyperlink{q:mdr}{MDR}                     & \SI{100}{TV}                               & for $|Z|=1$                                      \\
    \hyperref[tab:magnet]{Material budget}     & $0.12\,X_0$                                &                                                  \\
    of main solenoid                           & $0.012\,\lambda_I$                         &                                                  \\
    \hyperlink{q:calo}{Calorimeter depth}      & $70\,X_0$, $4\,\lambda_I$                  &                                                  \\
    Energy reach                               & \SI[retain-unity-mantissa=false]{1e16}{eV} & \hyperref[fig:protons]{for nucleons}             \\
                                               & \SI{10}{TeV}                               & for \hyperref[fig:positrons]{$e^+$}, \hyperref[fig:pbar]{$\bar{p}$}   \\
                                               & \SI{8}{GeV/n}                              & \hyperref[fig:dbar]{for $\bar{D}$}               \\
    \hyperref[fig:photons]{Angular resolution} & \ang{;;4}                                  & for photons at \SI{1}{TeV}                       \\
                                               & \ang[angle-symbol-over-decimal]{;;0.4}     & for photons at \SI{10}{TeV}                      \\
    \hyperlink{q:scifi}{Spatial resolution (SciFi)}      & \SI{40}{\um}                     &                                                  \\
    \hyperlink{q:si}{Spatial resoultion (Si-Tracker)}    &  \SI{5}{\um}                     &                                                  \\
    \hyperlink{q:tof}{Time resolution of single ToF bar} & \SI{20}{ps}                      &                                                  \\
    \hyperlink{q:rawrate}{Incoming particle rate}        & \SI{2}{MHz}                      &                                                  \\
    \hyperlink{q:trig}{High-level trigger rate}          & few kHz                          &                                                  \\
    \hyperlink{q:datarate}{Downlink data rate}           & $\sim$\SI{28}{Mbps}              &                                                  \\
    \hyperref[tab:weight]{Instrument weight}             & \SI{43}{\tonne}                  &                                                  \\
    \hyperlink{q:chan}{Number of readout channels}       & 8 million                        &                                                  \\
    \hyperlink{q:power}{Power consumption}               & \SI{10}{kW}                      &                                                  \\
    \hyperlink{q:duration}{Mission flight time}          & 10~years                         &                                                  \\
  \end{tabular}
  \caption[AMS-100 quantities.]{Important quantities in the AMS-100
    design.}
  \label{tab:quantities}
\end{table*}

\subsection*{Acknowledgments}
We thank the ACE/MAG instrument team and the ACE Science Center for
providing the ACE data.

\section*{References}

\bibliography{DetectorConcept}

\end{document}